\documentclass{emulateapj}

\usepackage{array} 
\usepackage{amsfonts}
\usepackage{amssymb} 
\usepackage{amsmath}
\usepackage{graphicx}
\usepackage{enumerate}
\usepackage{verbatim}
\usepackage{color}
\usepackage{mathrsfs}
\usepackage{comment}
\usepackage{multirow} 
\usepackage{diagbox}
\bibpunct{(}{)}{;}{a}{}{,} 
\usepackage{url}
\usepackage{hyperref}
\usepackage{breakurl}

\usepackage{verbatim}

\newcommand{\cmlt}{$\alpha_{\rm MLT}$} 

\newcommand{\rhds}{3D RHD simulations} 

\newcommand{\ssim}{s_{\rm ad, sim.}} 
\newcommand{\sint}{s_{\rm ad, int.}}

\shorttitle{Solar models with entropy-based radius calibration}
\shortauthors{Spada et al.}
\begin{document}


\title{Improved calibration of the radii of cool stars based on 3D simulations of convection: implications for the solar model}


\author{F. Spada}
\affiliation{Max-Planck Institut f\"ur Sonnensystemforschung, Justus-von-Liebig Weg 3, G\"ottingen, Germany}

\author{P. Demarque}
\affiliation{Department of Astronomy, Yale University, New Haven, CT 06520-8101, USA}

\author{S. Basu}
\affiliation{Department of Astronomy, Yale University, New Haven, CT 06520-8101, USA}

\author{J. D. Tanner}
\affiliation{Department of Astronomy, Yale University, New Haven, CT 06520-8101, USA}

\email{spada@mps.mpg.de}

\begin{abstract}
Main sequence, solar-like stars ($M \lesssim 1.5 \, M_\odot$) have outer convective envelopes that are sufficiently thick to affect significantly their overall structure.
The radii of these stars, in particular, are sensitive to the details of inefficient, super-adiabatic convection occurring in their outermost layers. 
The standard treatment of convection in stellar evolution models, based on the Mixing-Length Theory (MLT), provides only a very approximate description of convection in the super-adiabatic regime. 
Moreover, it contains a free parameter, \cmlt{}, whose standard calibration is based on the Sun, and is routinely applied to other stars ignoring the differences in their global parameters (e.g., effective temperature, gravity, chemical composition) and previous evolutionary history.

In this paper, we present a calibration of \cmlt{} based on three-dimensional radiation-hydrodynamics (3D RHD) simulations of convection.
The value of \cmlt{} is adjusted to match the specific entropy in the deep, adiabatic layers of the convective envelope to the corresponding value obtained from the 3D RHD simulations, as a function of the position of the star in the $(\log g, \log T_{\rm eff})$ plane and its chemical composition.

We have constructed a model of the present-day Sun using such entropy-based calibration. 
We find that its past luminosity evolution is not affected by the entropy calibration. The predicted solar radius, however, exceeds that of the standard model during the past several billion years, resulting in a lower surface temperature.
This illustrative calculation also demonstrates the viability of the entropy approach for calibrating the radii of other late-type stars.
\end{abstract}

\def\aj{{AJ}}                   
\def\araa{{ARA\&A}}             
\def\apj{{ApJ}}                 
\def\apjl{{ApJ}}                
\def\apjs{{ApJS}}               
\def\aap{{A\&A}}                
\def\apss{{Ap\&SS}}          
\def\aapr{{A\&A~Rev.}}          
\def\aaps{{A\&AS}}              
\def\mnras{{MNRAS}}             
\def\nat{{Nature}}              
\def\ssr{{Space~Sci.~Rev.}}

\section{Introduction}
\label{introduction}

The treatment of convection in stellar envelopes is one of the largest sources of uncertainty in the modeling of late-type stars (mid-F to later spectral types; $M \lesssim 1.5 \, M_\odot$).
Their outer layers are characterized by the presence of a thick convective envelope due to the very high value of the opacity in these regions \citep[see, e.g.,][]{S58,C68,KWW12}.
The result is a complex interaction in the energy transfer between radiative processes and relatively inefficient convective transport, particularly in the superadiabatic transition layer (SAL), located between the radiative photosphere and the regions where convection is efficient. 

Primarily because of its convenience and ease of implementation, the mixing-length theory \citep[MLT;][]{BV58} is usually the standard approach employed in the calculation of the convective temperature gradient and the structure of the SAL in stellar evolution codes.
The MLT approach has, however, several drawbacks \citep[see, e.g.,][]{Trampedach:2010},
one of the most severe being the introduction of a free parameter, the mixing-length $\ell$, that is completely unconstrained by the theory.
This limitation is shared with the other one-dimensional descriptions of convection alternative to the MLT that are available in the literature \citep[e.g.,][]{Canuto_Mazzitelli:1991,Arnett_ea:2010}.

The MLT parameter can be interpreted as a measure of convective efficiency, and it is most commonly expressed in a non-dimensional form as $\alpha_{\rm MLT}=\ell/H_P$, i.e., as the mixing length $\ell$ divided by the local pressure scale height $H_P$.
Although the MLT contains several other free parameters (see, e.g., the discussions of the MLT formalism by \citealt{Tassoul_ea:1990,Ludwig_ea:1999,Salaris_Cassisi:2008,Arnett_ea:2010,Tanner_ea:2016}), these are usually kept fixed in standard applications, and the major uncertainty is encompassed within the choice of \cmlt{}.  

The freely adjustable scale factor \cmlt{} sets the constant value to which the specific entropy asymptotically converges when moving from the SAL towards the deeper convective layers, where convection is efficient and the stratification nearly adiabatic.
Since, for a given metallicity, the radius of a star and the depth of its outer convection zone are both determined by the value of the adiabatic specific entropy, $s_{\rm ad}$ \citep[see, e.g.,][]{Stahler:1988}, they are both functions of the choice of \cmlt{}.
In other words, a wide range of interior structures is permitted as a function of \cmlt{}, and, as a result, the radius scale of stellar models is essentially freely adjustable, and requires a calibration external to the MLT.

Indeed, with MLT models alone, there is no way to determine which asymptotic entropy, or which adiabat, is the correct one for a given stellar model.  
As an illustration, the reader is referred to the upper panel of figure 1 of \citet{Tanner_ea:2016}, which shows the specific entropy profiles of four 1D stellar models with identical stellar atmospheric parameters, each computed with a different value of \cmlt{}.   

Except for the relatively thin region of the SAL, the superadiabaticity of the temperature gradient required to carry out the energy flux in the convection zone is of the order of one part in a million \citep[see, e.g.,][]{S58,KWW12}.  
As a result, the structure in the deep convective layers is very close to being polytropic, with a scale factor set by its specific entropy. 
All the specific entropy profiles shown in figure 1 of \citet{Tanner_ea:2016} share the qualitative features described above, but are characterized by a different value of $s_{\rm ad}$.

Besides the uncertainty introduced by the choice of \cmlt{}, another issue associated with the use of MLT in stellar evolution calculations is that the parameter \cmlt{} is typically taken to be a constant for all models along an evolutionary track, irrespective of their position in the $\log g$-$\log T_{\rm eff}$ plane or their chemical composition.  
The dependence of the properties of convection on the stellar parameters that change during the evolution, such as surface gravity, effective temperature, and surface metallicity, is therefore ignored.   
This is the case in spite of mounting evidence suggesting that \cmlt{} should depend on the metallicity of the star and its location in the HR diagram \citep{Lebreton_ea:2001, Yildiz_ea:2006,Bonaca_ea:2012,Viani_ea:2018}.
\citet{Tayar_ea:2017} found that a metallicity-dependent value of \cmlt{} is required to reconcile the stellar model predictions with the observed parameters of red giants stars from the APOGEE-{\it Kepler} catalog. 
This result, however, depends significantly on the choice of the outer boundary conditions employed in the models \citep{Salaris_ea:2018}.

Furthermore, both helioseismic observations and three-dimensional radiation hydrodynamics (3D RHD) simulations reveal that the MLT does not model faithfully the structure and dynamics of the non-adiabatic outer convective layers.  
This is the case of the SAL, which in turn controls the extent of the entropy jump (i.e., the difference between the adiabatic and the photospheric specific entropy), the depth of the convection zone, and the radius of the star.

Traditionally, in the calculation of stellar evolution grids, the value of \cmlt{} has been calibrated on a Standard Solar Model \citep[SSM; see, e.g.,][]{Basu_Antia:2008}. 
The Sun is the star for which we have the most reliable information, and unique independent knowledge of the mass, radius, and age. 
The main constraints of the SSM are the observed parameters of the Sun: its mass and radius, in addition to its age (from meteoritic ages), and its atmospheric chemical composition parameter $Z/X$ (where $Z$ and $X$ are the mass fractions of metals and hydrogen, respectively).  
In order to obtain a model of the Sun that fits these constraints, three parameters are adjusted: the initial metal fraction, $(Z/X)_0$, the initial helium, $Y_0$, and \cmlt{}. 
Roughly speaking, these three parameters mostly affect the solar present-day $Z/X$, the luminosity, and the radius, respectively.
A SSM calibration thus provides a handle on the choice of \cmlt{}.

In view of the -- perhaps surprisingly -- satisfactory results of this prescription, modeling convection using the MLT and the solar-calibrated value of \cmlt{}, as well as keeping this value constant along the stellar evolution track, has become the standard approach (for instance, this procedure can reproduce the effective temperature scale of the red giant branch of Galactic globular clusters: see \citealt{Salaris_ea:2002}, and references therein).
However, since \cmlt{} determines the properties of convection in the star, and these depend, in turn, on its parameters, using the solar-calibrated value of \cmlt{} to model stars that are very far from the solar-like regime (e.g., at different evolutionary stages, or having different chemical composition) is clearly unjustified. 

The limitations of the mixing length approximation have led to studies of stellar convection using 3D RHD numerical simulations. 
Simulations have been applied to dwarf stars \citep[e.g.,][]{Ramirez_ea:2009}, giants \citep[e.g.,][]{Ludwig_Kucinskas:2012}, and several targeted studies of individual stars \citep[e.g.,][]{Robinson_ea:2004, Robinson_ea:2005, Straka_ea:2007, Ludwig_ea:2009, Behara_ea:2010}.

Efforts to systematically study the variation of stellar convection have been carried out by, e.g., \citet{Ludwig_ea:1995, Ludwig_ea:1998, Ludwig_ea:1999, Freytag_ea:1999, Trampedach_Stein:2011, Tanner_ea:2013a, Tanner_ea:2013b, Tanner_ea:2014, Magic_ea:2013a, Trampedach_ea:2013}. 
Much interest is currently focused on determining how the properties of convection derived from 3D RHD simulations can be applied to 1D models of stars.  

\citet{Salaris_Cassisi:2015} have implemented in a 1D stellar evolution code the atmospheric temperature stratification and calibrated \cmlt{} (dependent on the effective temperature and surface gravity) from the 3D RHD simulations of \citet{Trampedach_ea:2014a,Trampedach_ea:2014b}.
Their calculations were performed for stars of mass between $0.75$ and $3.0\, M_\odot$ within the range of $\log g$--$\log T_{\rm eff}$ covered by the 3D simulations.
The models show only modest differences (e.g., $\lesssim 50$ K in $T_{\rm eff}$) with respect to their constant, solar-calibrated \cmlt{} counterparts.

An independent analysis based on the same 3D RHD calibrations \citep{Trampedach_ea:2014a,Trampedach_ea:2014b}, but differing in some key technical aspects and general goal, was performed by \citet{Mosumgaard_ea:2018} (see their paper for details).

Another promising approach was recently outlined by \citet{Tanner_ea:2016}, who suggested to use the value of the specific adiabatic entropy, $s_{\rm ad}$, extracted from the 3D RHD simulations, as the key parameter for the calibration of \cmlt{} in 1D stellar evolution codes.
These authors also showed that $s_{\rm ad}$ can be conveniently expressed as a function of a single variable defined through an opportune projection in the $\log g$--$\log T_{\rm eff}$ plane.

In this paper, we address the problem of stellar radii in the presence of a convection zone using an entropy-based approach to the calibration of \cmlt{}, as proposed by \citet{Tanner_ea:2016}.  
This work has two main objectives, as described below.

The first objective is to describe the implementation into a stellar evolution code of an improved calibration of the MLT parameter, alternative to the solar-based approach. 
Our calibration is based on the evaluation of the adiabatic specific entropy $s_{\rm ad}$ obtained from 3D RHD simulations as a function of chemical composition and position in the $(\log g, \log T_{\rm eff})$ plane \citep{Tanner_ea:2016}.  
The numerical value of \cmlt{} is then adjusted in the 1D interior model at each evolutionary time step, so as to yield the same value of $s_{\rm ad}$ obtained in the 3D RHD simulations (we use the results of \citealt{Magic_ea:2013a,Magic_ea:2013b,Magic_ea:2015a,Magic_ea:2015b} and of \citealt{Tanner_ea:2013a,Tanner_ea:2013b,Tanner_ea:2014}, as distilled in the fitting formulae provided by \citealt{Tanner_ea:2016}).  
The method continues to rely on the MLT formalism to construct the structure of the convective envelope, and thus determine $s_{\rm ad}$ as a function of the value of \cmlt{} adopted at each time step. 
This effective \cmlt{}, however, is variable along the evolutionary track.
This procedure puts radius estimates on a more physical basis and removes the role of free parameters.
Its main advantages are the simplicity of its implementation in an existing stellar evolution code, and the shift from the code-dependent, purely numerical parameter \cmlt{} to the physical variable $s_{\rm ad}$ as the fundamental quantity of the calibration.   
 
The second objective of this paper is to illustrate the effect of the improved radius calibration by exploring the past and future evolution of the Sun, and to compare the results with the conventional stellar evolution tracks constructed with a constant \cmlt{}.
The revised role of the SSM in stellar evolution calculations is briefly discussed. 

We note that a number of open issues of current astrophysics can essentially be ascribed to the need of a more reliable, physically motivated calibration of stellar radii; this paper is intended as a first step in that direction. 

For example, the measured radii of low-mass, main sequence stars are often found to be in disagreement with model predictions \citep{Hoxie:1973,Lacy:1977, Lopez-Morales:2007, Torres_ea:2010, Feiden_Chaboyer:2012a, Spada_Demarque:2012, Spada_ea:2013,Somers_Pinsonneault:2015, Lanzafame_ea:2017, Spada_ea:2017}. 
The effect of magnetic fields on convection is usually invoked to explain this discrepancy \citep[e.g.,][]{Chabrier_ea:2007, Feiden_Chaboyer:2012b, Feiden_Chaboyer:2013, Feiden_Chaboyer:2014, Feiden:2016}. 
An improvement in the standard treatment of convection would allow a refined assessment of such discrepancy, and provide a more robust starting point for the development of more advanced models. 

Furthermore, precise modeling of red giant stars, made possible by asteroseismology, seems to require the introduction of a metallicity-dependent \cmlt{} \citep{Piau_ea:2011, Tayar_ea:2017, Li_ea:2018}. 

Finally, estimating the radii of exoplanets, which is essential to constrain their density and interior structure, relies on the accurate characterization of the host star \citep{Boyajian_ea:2015,Shields_ea:2016}.

This paper is organized as follows: in Section \ref{methods} we describe our stellar evolution code and the details of the entropy-based calibration procedure of the MLT parameter \cmlt{}. 
In Section \ref{ssm_vs_esm} we discuss the uncertainties associated with the construction of the SSM, and we describe our procedure to construct an entropy-calibrated solar model.
In Section \ref{results} we compare the interior properties and the evolution of an entropy-calibrated solar model with one constructed using the standard solar calibration procedure.
Our results are discussed in Section \ref{discussion}.
Our main conclusions are summarized in Section \ref{conclusions}.

\section{Methods}
\label{methods}

\subsection{An alternative calibration for stellar radii}

In our stellar evolution models, \cmlt{} is not calibrated against the solar radius, as in the standard approach, but by matching the entropy of the deep portion of their convective layers, where convection is in the adiabatic regime, with the respective value obtained from 3D RHD simulations \citep[see][]{Tanner_ea:2016}.
The parameter \cmlt{} is re-calibrated at each step in the evolutionary sequence, i.e., for the current values of metallicity, [Fe/H], effective temperature, $T_{\rm eff}$, and surface gravity, $\log g$, at the surface of the star.  
This procedure preserves the dependence of the properties of convection on the stellar atmospheric parameters, as distilled in the results of the 3D RHD simulations. 
In turn, it provides an alternative calibration of the surface radius for stellar models.

\begin{figure}
\begin{center}
\includegraphics[width=0.49\textwidth]{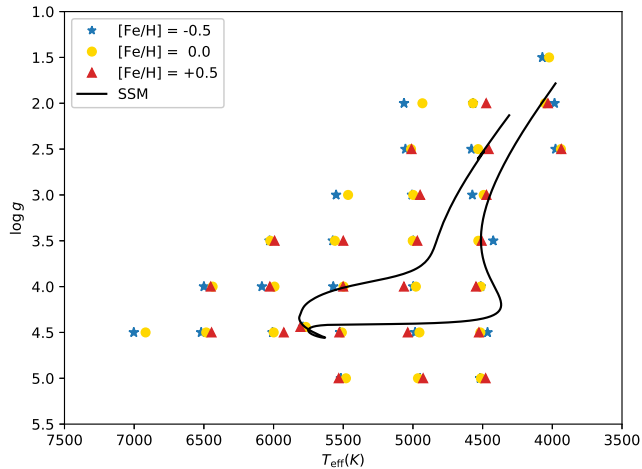}
\caption{Coverage of the $(\log T_{\rm eff}, \log g)$ plane by the \texttt{Stagger} grid simulations, together with the evolutionary track of our SSM for comparison. Only the simulations in a range of metallicity close to solar are shown for clarity (cf. \url{https://staggergrid.files.wordpress.com/2014/08/stagger-grid.png}).}
\label{fig_stagger}
\end{center}
\end{figure}

The set of parameters $\left\{{\rm [Fe/H]}, T_{\rm eff}, \log g \right\}$ completely specifies the conditions of the stellar envelope, and therefore its adiabatic specific entropy, $s_{\rm ad}$. 
The \rhds{} can then be used to specify the mapping:
\begin{equation*}
\ssim{}({\rm [Fe/H]},T_{\rm eff}, \log g).
\end{equation*}
In this work, we rely on the mapping constructed by \citet{Tanner_ea:2016} based on the \rhds{} of both \citet{Tanner_ea:2013a,Tanner_ea:2013b,Tanner_ea:2014} and of the \texttt{Stagger} grid \citep{Magic_ea:2013a,Magic_ea:2013b,Magic_ea:2015a,Magic_ea:2015b}.
The larger of these two subsets of simulations, the \texttt{Stagger} grid, covers a range of $T_{\rm eff}$ from $4000$ to $7000$ K, $\log g$ from $1.5$ to $5.0$, and [Fe/H] from $-4.0$ to $+0.5$. 
The region of the HRD covered by the simulations at metallicity close to solar is illustrated in Figure \ref{fig_stagger}, together with the evolutionary track of our SSM for comparison.

Although these two subsets of simulations were constructed using independent codes (the \texttt{Stagger} code, Nordlund \& Galsgaard 1995\footnote{\url{www.astro.ku.dk/~kg/Papers/MHD_code.ps.gz}}, \citealt{Kritsuk_ea:2011}, and the CKS code, \citealt{Chan_Sofia:1989, Kim_Chan:1998}) which implement significantly different choices of input physics (e.g., equation of state, opacities, atmospheric structure, see the detailed comparison by \citealt{Kupka:2009}), \citet{Tanner_ea:2016} showed that the predicted values of $s_{\rm ad}$ are remarkably consistent with each other (see their figure 3). 

\citet{Tanner_ea:2016} provided the mapping of $\ssim$ in analytical form, as a function of the variable: 
\begin{equation}
\label{jdt_x}
x \equiv A\, \log T_{\rm eff} + B\, \log g.
\end{equation}
This transformation corresponds to a rotation in the $(T_{\rm eff}, \log g)$ plane (sometimes called the ``Kiel diagram"), and is motivated by the realization that the contours of constant $s_{\rm ad}$ are approximately straight lines in this plane (cf. figure 3 of \citealt{Ludwig_ea:1999}, and figure 2 of \citealt{Magic_ea:2015a} for 2D and 3D RHD simulations, respectively).
The functional form chosen by \citet{Tanner_ea:2016} is:
\begin{equation}
\label{jdt_fit}
\ssim{} = s_0 + \beta \exp{\left(\frac{x-x_0}{\tau}\right)}.
\end{equation}
The parameters $A$, $B$, $s_0$, $x_0$, $\beta$, $\tau$ are assigned as (metallicity-dependent) best-fitting coefficients to the 3D RHD simulations of \citet{Magic_ea:2015a}.

In this work, we have used the best-fitting coefficients given in table 1 of \citet{Tanner_ea:2016}. 
In order to test the robustness of the fit and the accuracy of equation \eqref{jdt_fit}, however, we have independently repeated the best-fitting procedure for the solar metallicity case; the results of such analysis are detailed in Appendix \ref{app_fits}.

The entropy of the adiabatic layers in a stellar interior model constructed with a 1D stellar evolution code is a function of the evolutionary stage, including the chemical composition, and of the MLT parameter \cmlt{}. 
Our entropy-based calibration of \cmlt{} therefore consists in choosing the value of this parameter for which the adiabatic entropy in the 1D stellar interior model, $\sint{}$, coincides with that obtained from the \rhds{}, $\ssim{}$:
\begin{equation}
\label{calibration}
\sint{} (\alpha_{\rm MLT}) \equiv \ssim{} \rightarrow \alpha_{\rm MLT}({\rm [Fe/H]},x),
\end{equation}
where the functional dependence of $\sint{}$ and $\ssim{}$ on [Fe/H], $T_{\rm eff}$, and $\log g$ were omitted for clarity. 

\begin{figure}
\begin{center}
\includegraphics[width=0.5\textwidth]{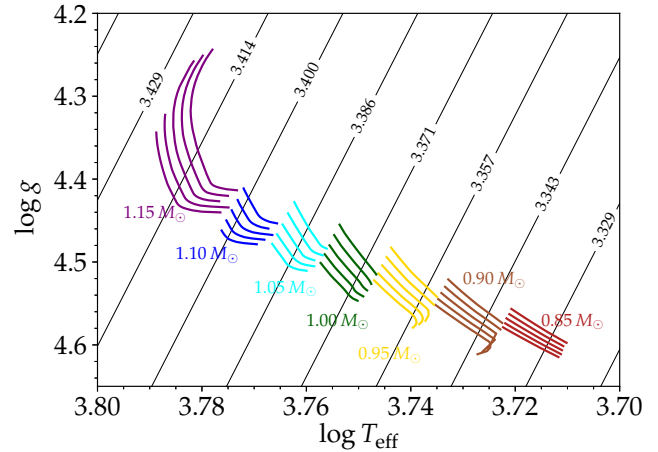}
\includegraphics[width=0.5\textwidth]{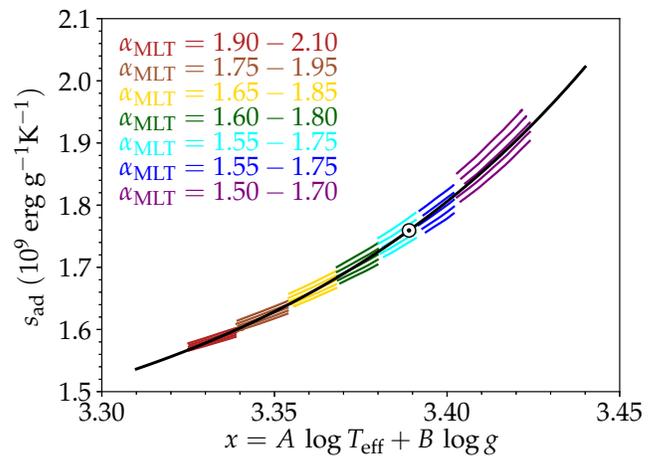}
\includegraphics[width=0.5\textwidth]{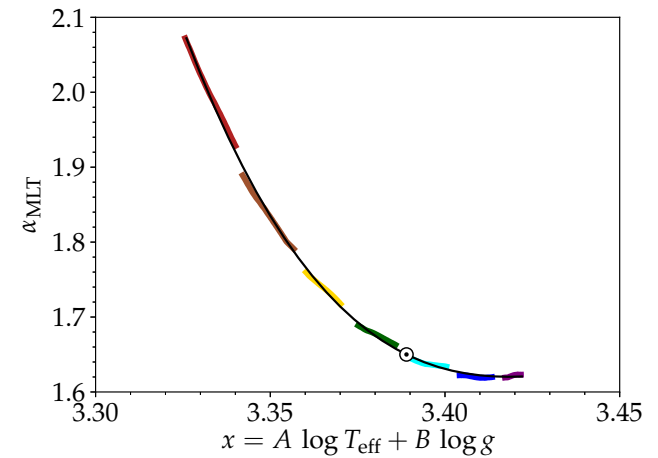}
\caption{Calibration of \cmlt{} against $s_{\rm ad}$.
Top panel: evolutionary tracks (colored lines) at solar metallicity in the Kiel diagram; contours of $x$ are shown in black.
Middle: $s_{\rm ad}$ vs. $x$ relation for the same tracks (colors) compared with equation (\ref{jdt_fit}; thick black line).  
Bottom: calibrated \cmlt{} obtained by interpolating the evolutionary tracks at fixed $x$; the black line shows a fit to the resulting $\alpha_{\rm MLT}(x)$.}
\label{fig:calibration_tracks}
\end{center}
\end{figure}

Figure \ref{fig:calibration_tracks} illustrates the basic concept of the calibration procedure.
The upper panel of the Figure shows a set of evolutionary tracks of solar composition and for different values of mass and \cmlt{}, constructed with a 1D stellar evolution code (see Section \ref{thecode}); lines of constant $x$ are also shown.
For each evolutionary track, a portion of the main sequence evolution is shown, between the ZAMS and the TAMS, or shorter.
For this section only, gravitational settling (or ``diffusion") of helium and heavy elements is neglected for simplicity, whereas it is taken into account in the models discussed in the rest of the paper.

The adiabatic specific entropy along the same tracks, plotted as a function of $x$, intersects the relation defined by equation \eqref{jdt_fit} at solar metallicity (middle panel).
For each mass, several values of \cmlt{} were used to construct a subset of tracks that locally bracket the $\ssim{}(x)$ relation.
At fixed $x$, the value of \cmlt{} that satisfies equation \eqref{calibration} can then be found by interpolating among those tracks.
The resulting calibrated \cmlt{} is plotted as a function of $x$ in the lower panel of Figure \ref{fig:calibration_tracks}.

From the middle panel of Figure \ref{fig:calibration_tracks}, it can be seen that all the tracks of the subset at $M=0.85\, M_\odot$ intersect the $\ssim{}(x)$ relation, whereas at $M=1.15\, M_\odot$ the track with $\alpha_{\rm MLT}=1.60$ approximately coincides with $\ssim{}(x)$ in the whole subrange of $x$ shown.
As a result, the slope of the curve $\alpha_{\rm MLT}(x)$ is steeper in the $x$ range covered by the low-mass tracks, and flatter in the $x$ range covered by tracks of higher mass.
The function $\alpha_{\rm MLT}(x)$ decreases monotonically from $2.1$ to $1.6$ as $x$ increases in the range $3.32$ to $3.42$.

For the Sun, $x_\odot = 3.389$, and $\alpha_{\rm MLT}(x_\odot)=1.650$. 
This value should be compared with the results of a standard solar calibration: $\alpha_{\rm MLT,\odot}=1.825$ if diffusion is taken into account, and $\alpha_{\rm MLT,\odot}=1.693$ if diffusion is ignored (cf. Section \ref{ssm_construction}).

This difference is due to a (moderate) inconsistency between the values of $s_{\rm ad}$ obtained from equation \eqref{jdt_fit} and from the SSM calibration. 
As discussed in Appendix \ref{app_fits}, this inconsistency is due to the original choice of the best-fitting parameters by \citet{Tanner_ea:2016}, and results in an entropy-calibrated solar model with an incorrect radius (by $\approx 1\%$).
To address this issue, we have introduced a correction in equation \eqref{jdt_fit} to compensate for this offset in $s_{\rm ad}$ (see Section \ref{results} for details).

In the following, we present a revised evolutionary track and interior model for the Sun, calculated using the \cmlt{} calibration procedure just described.
In order to achieve self-consistency, and to include the effect of helium and metal diffusion, the calibration is performed at each evolutionary time step.
We have therefore developed a procedure to construct an evolutionary sequence of models that re-calibrates \cmlt{} after each converged model is obtained.

\begin{figure}
\begin{center}
\includegraphics[width=0.49\textwidth]{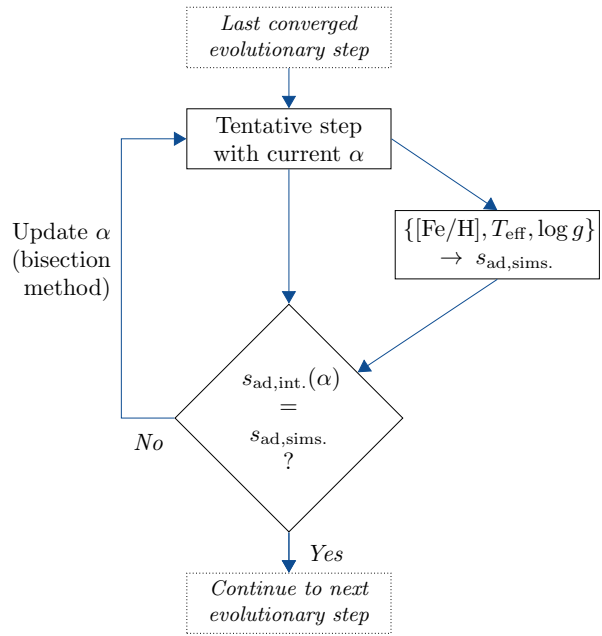}
\caption{Schematic of the \cmlt{} calibration loop included at each evolutionary step of our evolutionary calculations.}
\label{flowchart}
\end{center}
\end{figure}

\subsection{Implementation in the stellar evolution calculation}
\label{thecode}

The models discussed in this paper were constructed using the Yale Rotational stellar Evolution Code (YREC) in its non-rotational configuration \citep{Demarque_ea:2008}.
The standard MLT description of convection is used in the code \citep{BV58}. 
Diffusion of helium and heavy elements is taken into account in the models discussed from now on, according to the formulation of \citet{Thoul_ea:1994}.
We adopted the OPAL 2005 equation of state \citep{Rogers_Nayfonov:2002}, the OPAL opacities \citep{Rogers_Iglesias:1995,Iglesias_Rogers:1996}, supplemented by the low-temperature opacities of \citet{Ferguson_ea:2005}; in the atmosphere, the Eddington gray $T$--$\tau$ relation is used.
We assume the \citet{Grevesse_Sauval:1998} solar mixture, and the corresponding value of $(Z/X)_\odot=0.0230$.

The code was used to construct evolutionary sequences in which each converged model is required to respect the constraint of having an adiabatic specific entropy in the deep convective envelope, $\sint{}$, that is consistent with $\ssim{}$ obtained from equation \eqref{jdt_fit} evaluated at the current values of $\left\{ {\rm [Fe/H]}, T_{\rm eff}, \log g \right\}$.
This is achieved by re-calibrating and adjusting the value of the mixing length parameter \cmlt{} when evolving from the latest converged model to the next.
A sketch of the calibration loop performed at each evolutionary time step is shown in Figure \ref{flowchart}.

The RHD-calibrated value of \cmlt{} is derived through a standard bisection root-finding procedure. 
Operationally, the bisection scheme runs YREC several times for each time step, until equation \eqref{calibration} is satisfied within a specified tolerance. 
For a tolerance of $10^{-4}$, the bisection loop converges within $\lesssim 10$ steps. 
Both the bisection procedure and a wrapper routine that handles the calling of YREC were written in the Python programming language.

\section{Standard vs. entropy-calibrated solar models}
\label{ssm_vs_esm}

\subsection{The SSM: concept and uncertainties}
\label{ssm_discussion}

The SSM has played a major role in testing the physics undelying the theory of stellar evolution, and the fundamental physics of neutrino flavor oscillations.  
The remarkable agreement of the helioseismic tests of the solar interior, which led to a clarification of the solar neutrino problem, and the realization that the problem is due to the failure of the Standard Model of particle physics and the incompleteness of neutrino theory, provides a test of the structure of the central region of the Sun.
More recently, detailed analysis of the uncertainties in the SSM assumptions focus on the solar chemical composition and the related input physics, such as the opacities \citep[e.g.,][]{Vinyoles_ea:2017,Trampedach:2018}. 
These composition uncertainties affect a wide variety of astrophysical problems in which the solar abundance is used as a benchmark, such as studies of galactic stellar populations. 

Before we compare the properties of the SSM with the entropy-calibrated solar models we have constructed, one important point must be made.  
While the deep interior of the SSM has extraordinary agreement with observation (primarily through seismology and neutrino observations), the description of the outer layers of the solar model is much less secure.  
In particular, the radius calculation is subject to our use of the MLT.  
The freedom to adjust the value of \cmlt{} to reproduce the solar radius precisely has been an advantage in constructing the SSM, providing the opportunity to lump all uncertainties within a single parameter. 
At the same time, by providing a solar model with a radius that precisely matches the observed solar radius, the SSM approach has masked the level of uncertainty in modeling the outer layers of the Sun, which are made evident by the so-called surface effects in the helioseismic frequencies.

\begin{table}
\caption{Parameters of the SSM, generated with YREC using the standard approach, i.e., the value of \cmlt{} is kept constant along the evolutionary track.}
\begin{center}
\begin{tabular}{ccc}
\hline
\hline
Parameter & Adopted & Model \\
\hline
Age (Gyr) & $4.57$ & (Exact) \\
Mass (g) & $1.9891\cdot 10^{33}$ & (Exact) \\
Radius (cm) & $6.9598 \cdot 10^{10}$ & $\log \frac{R}{R_\odot} =-1.7\cdot 10^{-7}$ \\
Luminosity (erg s$^{-1}$) & $3.8418\cdot 10^{33}$ &  $\log \frac{L}{L_\odot} = -5.4\cdot 10^{-6}$ \\
Surface $(Z/X)$ & 0.0230 & [Fe/H] $= 7.2 \cdot 10^{-5}$ \\
Surface $X$ & - &  0.735978399 \\
Surface $Z$ & - &  0.016930314 \\
\hline
\end{tabular}
\end{center}
\label{tab_ssm}
\end{table}

As examples, we point out two sources of uncertainties in constructing the SSM.  
One is on the external input of the age of the Sun.  
The other is related to the efficiency of the diffusion of helium and heavy elements in the solar interior, primarily due to the effects of rotationally driven turbulence \citep{Chaboyer_ea:1995a,Chaboyer_ea:1995b}.  
These uncertainties can affect the solar model radius (and the estimate of $s_{\rm ad}$) by several percent. 

The age of the Sun most frequently adopted for the SSM is $4.57$ Gyr, based on the lower limit set by the age of the oldest meteorites \citep{Baker_ea:2005}.  
This limit is to be compared to the lower limit of $4.6 \pm 0.05$ Gyr usually provided for the Earth.  
Until we fully understand the early evolution of the Sun in conjunction with the planetary system formation chronology, there remains an uncertainty of the order of $1\%$ in this input parameter.

We know from helioseismology that the process of diffusion of helium and heavy elements must be taken into account in the models in order to match the solar observations \citep{Christensen-Dalsgaard_ea:1993}.  
Diffusion is now recognized as an intrinsic part of the SSM construction. 
A detailed discussion of the combined effects of diffusion and rotational mixing was given by \citet{Chaboyer_ea:1995a,Chaboyer_ea:1995b}, with emphasis on $^7$Li depletion at the surface and on seismology.  

These, and the still not well understood structure of the base of the convection zone, present the largest current uncertainties in our understanding of the solar interior physics (see the recent paper on the B16 SSM model by \citealt{Vinyoles_ea:2017}).

\subsection{Construction of the SSM}
\label{ssm_construction}

An automatic calibration routine to construct a SSM is available as part of the standard YREC release \citep[see][]{Demarque_ea:2008}.
The code adjusts iteratively the values of the two initial composition parameters, $X_0$ and $Z_0$, and of the MLT parameter, \cmlt{}, to obtain a $1\, M_\odot$ evolutionary track having a radius equal to $R_\odot$, a luminosity equal to $L_\odot$, and $(Z/X)=(Z/X)_\odot=0.0230$ \citep{Grevesse_Sauval:1998} at solar age ($t_\odot = 4.57$ Gyr).

The evolution is initialized from a pre-computed early pre-main sequence model, with a simple polytropic structure (of index $N=1.5$) and homogenous composition.
For internal consistency and homogeneity with the rest of the computations presented in this paper, our reference SSM was constructed using a Python script driver that runs YREC  keeping \cmlt{} fixed during the evolution. 

With the choices of input physics detailed in Section \ref{thecode}, the calibrated values of the adjustable parameters are the following: $   X_{\rm 0, SSM} = 0.7038$; $Z_{\rm 0,SSM} = 0.01882$; $\alpha_{\rm MLT, SSM}=1.825$.
The fundamental parameters of the SSM are reported in Table \ref{tab_ssm}.

\begin{figure}
\begin{center}
\includegraphics[width=0.5\textwidth]{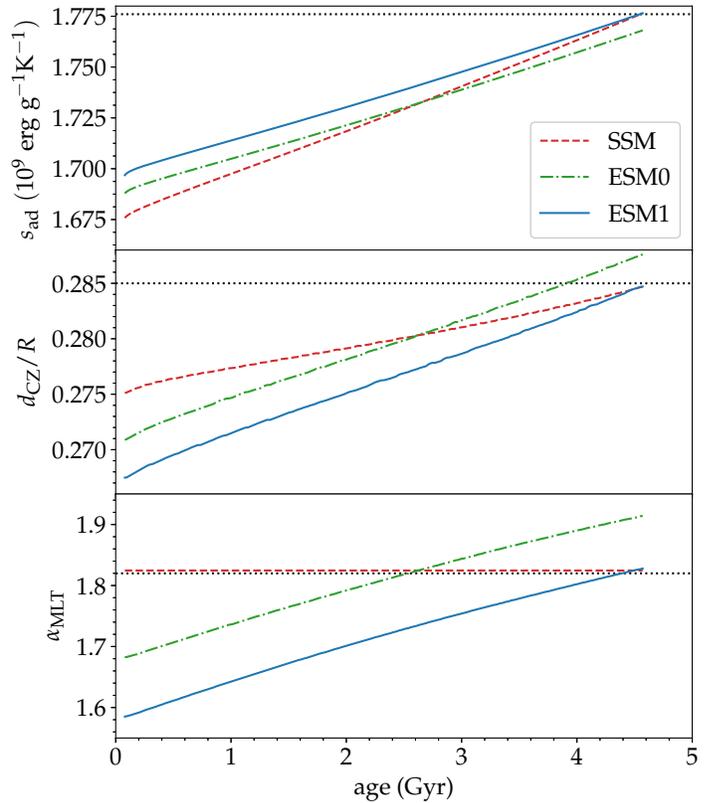}
\caption{Evolution from ZAMS to solar age of the specific entropy, the depth of the convection zone, and the MLT parameter for models SSM, ESM0, and ESM1. Note that model ESM1 has a shallower convective envelope than the SSM during the main sequence evolution until the present solar age is reached.}
\label{entropy}
\end{center}
\end{figure}

\subsection{Construction of the Entropy-calibrated Solar Model (ESM)} 

The properties of the ESM, and in particular its radius, depend critically on the exact value\footnote{Entropy is a state function and can always be redefined including an arbitrary additive constant. In the following, consistently with \citet{Tanner_ea:2016}, the values quoted for the adiabatic specific entropy are those obtained from the OPAL 2005 equation of state \citep{Rogers_Nayfonov:2002}.} of $\ssim{}$ against which \cmlt{} is calibrated along the evolution.
In particular, an inconsistency between the value of $s_{\rm ad}$ given by equation \eqref{jdt_fit} and that of the SSM unavoidably results in an inconsistent radius at solar age between the ESM and the SSM.

This is the case if equation \eqref{jdt_fit} and the values of the fitting parameters are adopted unmodified as given by \citet{Tanner_ea:2016}.
Indeed, equation \eqref{jdt_fit} evaluated at the solar parameters gives $\ssim{} = 1.760 \cdot 10^{9} \; {\rm erg}\, {\rm g}^{-1}\, {\rm K}^{-1}$, and it is therefore inconsistent with the SSM model discussed in the previous section, which has $s_{\rm ad}^{\rm SSM} = 1.776 \cdot 10^{9} \; {\rm erg}\, {\rm g}^{-1}\, {\rm K}^{-1}$.
The resulting entropy-calibrated model, referred to as ``ESM0" in the following, has $R(t_\odot) = 0.9892\, R_\odot$, and thus a radius discrepancy of approximately $1\%$, and a specific entropy difference of $\delta s_{\rm ad} \approx -0.9 \cdot 10^7$ with respect to the SSM. 

\begin{figure*}
\begin{center}
\includegraphics[width=0.49\textwidth]{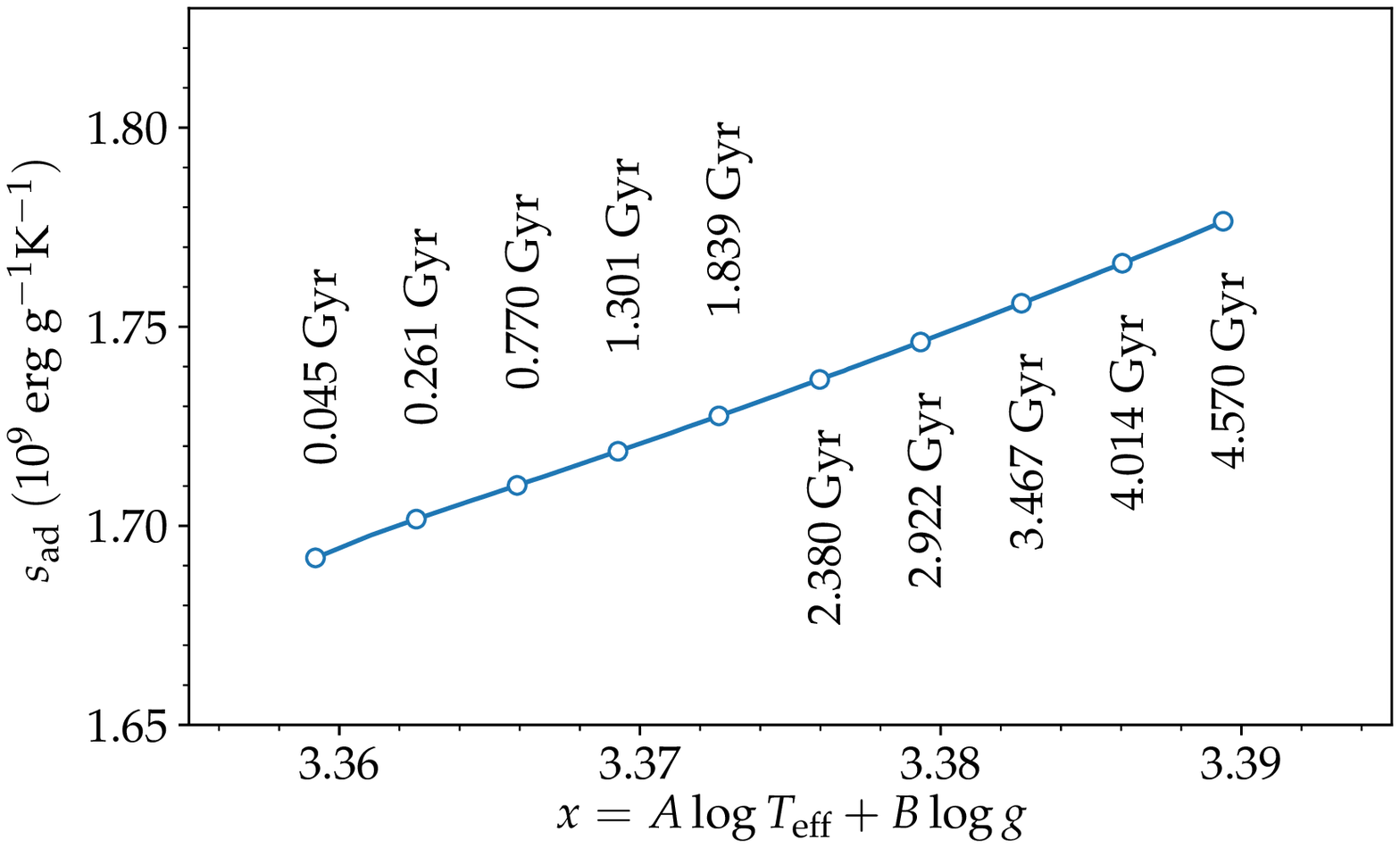}
\includegraphics[width=0.49\textwidth]{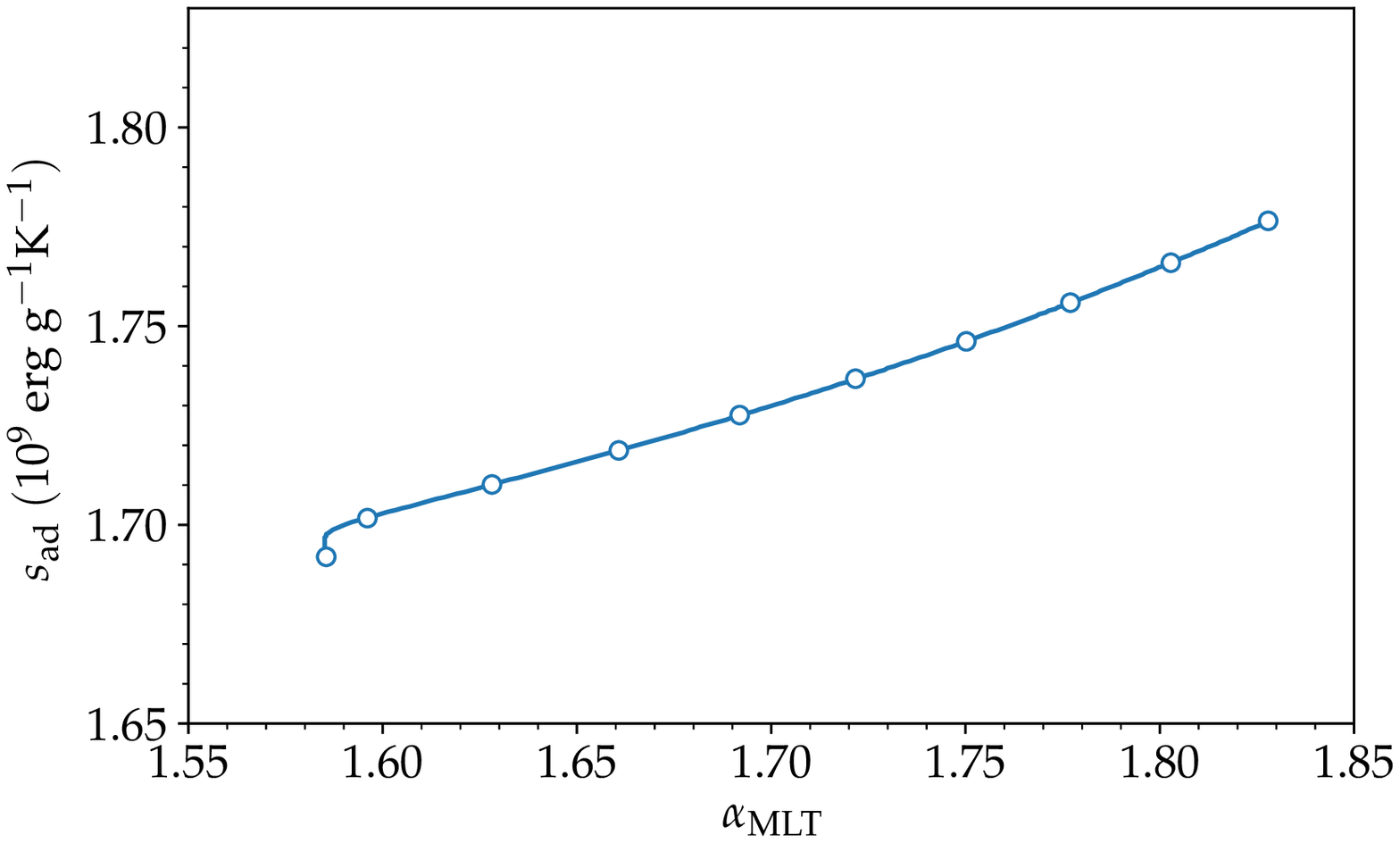}
\caption{Evolution from ZAMS to solar age of model ESM1 in terms of the main variables used in the calibration: $s_{\rm ad}$ vs. $x$ (left panel); $s_{\rm ad}$ vs. \cmlt{} (right panel). Ages at some representative points are reported in the labels.}
\label{entropy_plane}
\end{center}
\end{figure*}

It is possible to obtain a rough analytical estimate of the radius difference resulting from a given adiabatic specific entropy mismatch.
Assuming that the adiabatic layers of the solar convection zone behave as an ideal gas, and ignoring partial ionization, the following approximate relation holds (cf. equation 17 of \citealt{Ireland_Browning:2018}; see also \citealt{HKT04}):
$R \propto \exp \left( \mu s_{\rm ad}/ N_A k_B \right)$.
For model ESM0 we thus have:
\begin{equation*}
\frac{\delta R}{R} \sim \frac{\mu}{N_A k_B} \delta s_{\rm ad} \approx \frac{0.6}{8.3 \cdot 10^7} (-0.9 \cdot 10^7) \approx -6\%,
\end{equation*}
which agrees in sign with, but is somewhat larger than, the actual result quoted above. 
The depth of the convection zone in model ESM0 also differs from that of our SSM by approximately $1\%$.

Regarding the discrepancies between model ESM0 and the SSM, two important points must be emphasized.
First, the entropy-based model, in contrast with the classical SSM and its freely adjusted parameter \cmlt{}, sets a stringent constraint on the surface layers of the solar model.  
It is indeed encouraging that, despite the uncertainties still remaining in solar modeling, the mismatch in the independently evaluated values of $s_{\rm ad}$ at the interface between the interior model and the atmospheric 3D RHD simulations is limited to $1 \%$, and that it results in a discrepancy of $1 \%$ in the solar model radius.
It should also be noted that an offset of $1\%$ in $s_{\rm ad}$ with respect to the 3D RHD simulations, as found with the YREC code, is well within the combined uncertainty in the values of  $s_{\rm ad}$ associated with the simulations and the fitting procedure used by \citet{Tanner_ea:2016} to construct the fits in equation \eqref{jdt_fit}.

To deal with the entropy mismatch discussed so far, we have introduced an appropriate zero-point correction to the specific entropy $\ssim{}$ given by equation \eqref{jdt_fit}.
This offset, which amounts to $\delta s_{\rm ad,0} = 0.01585$, is taken into account during the evolution in the construction of our revised entropy-calibrated model, ``ESM1". 
In this way we recover the correct solar radius at $4.57$ Gyr in model ESM1. 

Given the very tight link between $s_{\rm ad}$ and $d_{\rm CZ}$, the correction introduced in model ESM1 can be thought of a calibration with respect to the depth of the convection zone of the present Sun, whose value is known with very high precision from helioseismology \citep{Christensen-Dalsgaard_ea:1991,Basu_Antia:1997}.

Although this correction is a form of calibration against the observed properties of the Sun, the time evolution of $\alpha_{\rm MLT}$ is entirely independent of $\delta s_{\rm ad,0}$, except for a vertical shift of its absolute value.
This latter point is illustrated by the evolution of the adiabatic specific entropy, the depth of the convection zone, and the MLT parameter, plotted in Figure \ref{entropy}.
The evolution of $s_{\rm ad}$ and $d_{\rm CZ}$  (upper and middle panel of Figure \ref{entropy}, respectively) of models ESM0 and ESM1 proceeds with an approximately constant offset. 
In other words, the correction introduced in model ESM1 is purely a shift of the zero point of the adiabatic specific entropy used in the calibration, which ensures that the value of $s_{\rm ad}$ at $4.57$ Gyr in model ESM1 coincides with that of the SSM.
This is also reflected in the evolution of the calibrated \cmlt{}, which attains the solar-calibrated value at solar age.

To further illustrate the calibration procedure, the evolution of the entropy-calibrated model in terms of the variables $x$, $s_{\rm ad}$, and \cmlt{} is plotted in Figure \ref{entropy_plane} (cf. Figure 4 of \citealt{Tanner_ea:2016}).
The entropy-calibrated \cmlt{} varies in the range $\approx 1.6$--$1.8$ during the evolution from the zero-age main sequence (ZAMS) to the solar age.

It should be emphasized that a different choice of the microphysics in the model will produce different values of the calibrated \cmlt{}.
The radius calibration, however, is only sensitive to the relative variation of \cmlt{} as a function of $x$ (see, e.g., the discussion of the impact of the $T$--$\tau$ relation in Section \ref{discussion}).

Since the MLT parameter can be interpreted as a measure of the efficiency of convection, smaller values of \cmlt{} result in a larger radius of the stellar model to compensate for the reduction in the efficiency of the energy transport.
The depth of the outer convection zone, and thus the its mixing properties and the surface composition, are also affected.

\section{Results}
\label{results}

\subsection{Interior structure of the ESM}
\label{interior}

In comparing the interior structure of model ESM1 with the SSM, we note, first of all, that their values of $s_{\rm ad}$, and thus \cmlt{}, are the same by construction.   
Moreover, we can expect the entropy calibration to have a very small impact on the evolution of the radiative interior and of the inner core.
On the contrary, the significant difference in the history of the depth of the convective envelope (cf. the middle panel of Figure \ref{entropy}) will result in differences in the chemical composition profile.
  
\begin{figure}
\begin{center}
\includegraphics[width=0.5\textwidth]{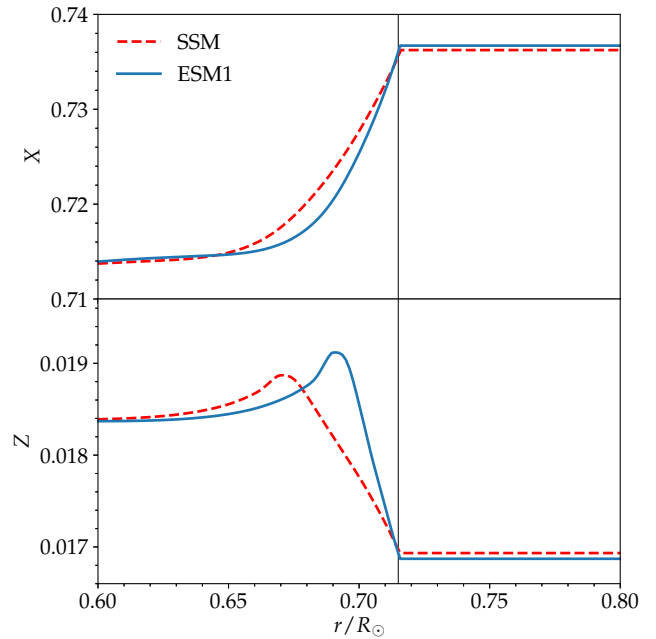}
\caption{Composition profiles in models ESM1 and in the SSM.}
\label{interior_xz}
\end{center}
\end{figure}

These expectations are confirmed by the plots of the the composition profile shown in Figure \ref{interior_xz}.

During the approximately $4.5$ Gyr of evolution between the ZAMS and the current solar age, the chemical composition of the inner core changes at the slow pace set by the timescale of nuclear reactions as hydrogen is converted into helium; in the radiative zone helium and heavy elements sink towards the core by gravitational settling, while the convection zone is kept at nearly constant composition by the comparatively much faster mixing provided by convective motions.

As Figure \ref{entropy} shows, the entropy-calibrated model is characterized by a shallower convective envelope in comparison with the solar-calibrated one during the main sequence evolution until the solar age.
As a result, the chemical composition differences between model ESM1 and the SSM are concentrated immediately below the current lower boundary of the convection zone, i.e., in the region that has undergone a less extensive convective mixing in the entropy-calibrated model with respect to the solar-calibrated one.
Since this is just a redistribution effect, the piling up of heavy elements below the convection zone boundary is compensated by their slightly lower abundances in the convective envelope itself in model ESM1 (cf. the lower panel of Figure \ref{interior_xz}).

The moderate differences in the composition profile in the entropy-calibrated model are accompanied by structural differences of the same order, or smaller.
Such subtle differences could only possibly be detected by helioseismology.   
Indeed, as pointed out by \citet{Vinyoles_ea:2017} (see also the review by \citealt{Serenelli:2016}), helioseismology reveals that the main remaining discrepancy between the observed oscillation frequencies and those predicted by the SSM is located in the tachocline region.  
This observed perturbation in the speed of sound appear to be caused by two effects: (1) the composition effect described in Section \ref{ssm_discussion}, and (2) the presence of a subadiabatic layer located just below the CZ due to boundary dynamical effects \citep{Christensen-Dalsgaard_ea:2011}.  
Such a layer is predicted by recent 3D simulations \citep{Kitiashvili_ea:2016,Kapyla_ea:2017}.  

\begin{figure}
\begin{center}
\includegraphics[width=0.5\textwidth]{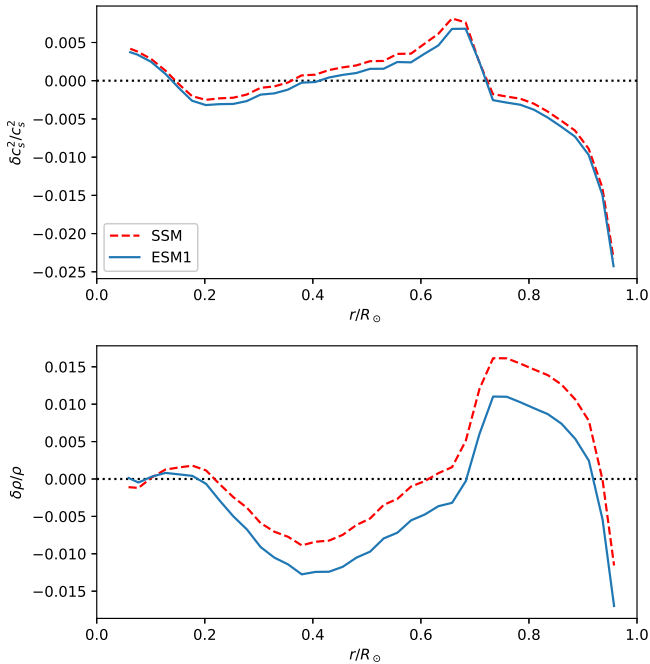}
\caption{Difference in the sound speed and density profiles with respect to the results of the BiSON-13 inversions (cf. table 3 of \citealt{Basu_ea:2009}) for models ESM1 and SSM.}
\label{interior_cs}
\end{center}
\end{figure}

The difference in sound speed and density with respect to the solar profile, as derived from the BiSON inversions by \citet{Basu_ea:2009}, are shown in Figure \ref{interior_cs}. 
The agreement of model ESM1 and of the SSM with the solar inversions is qualitatively similar.


\begin{figure*}
\begin{center}
\includegraphics[width=0.49\textwidth]{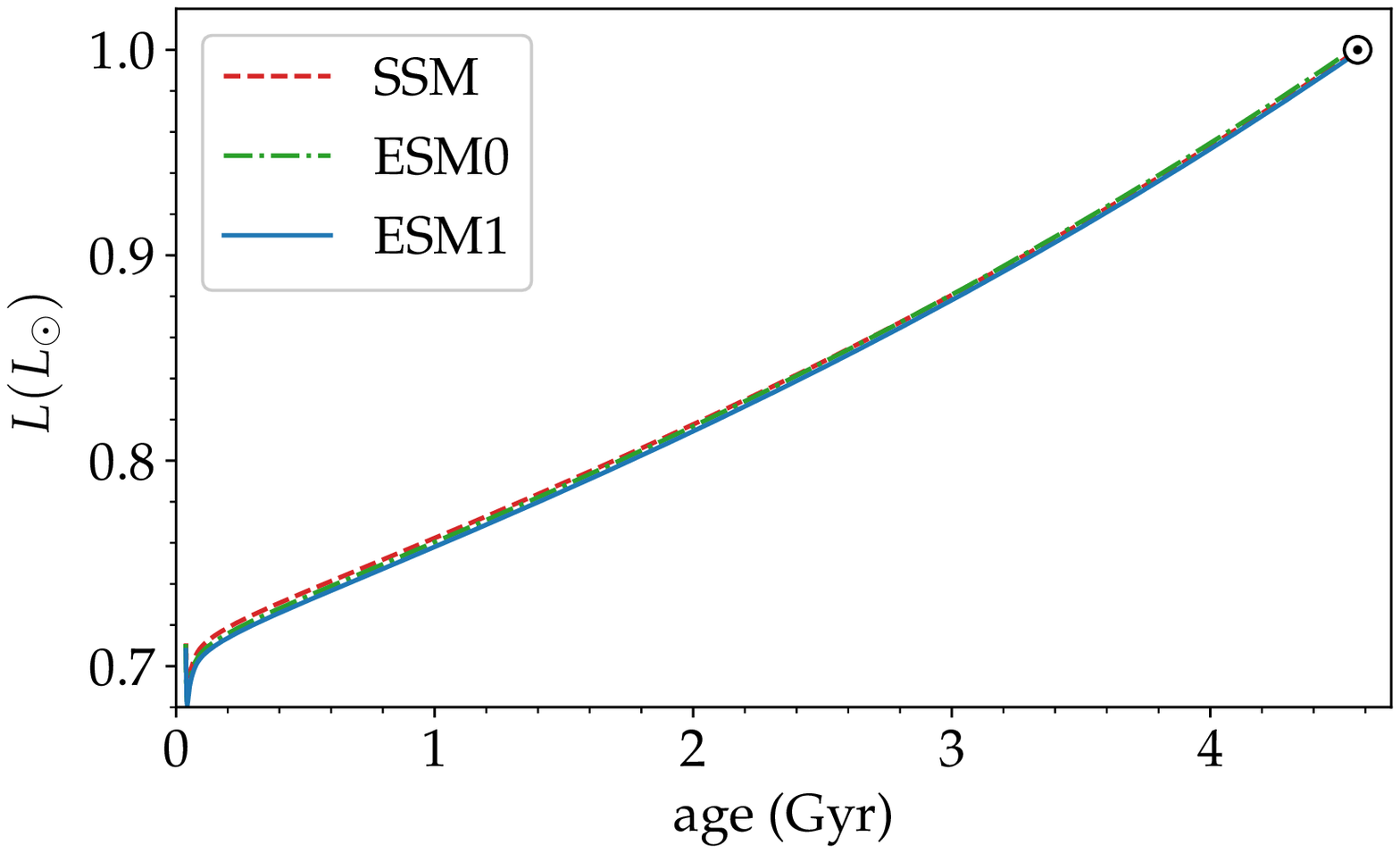}
\includegraphics[width=0.49\textwidth]{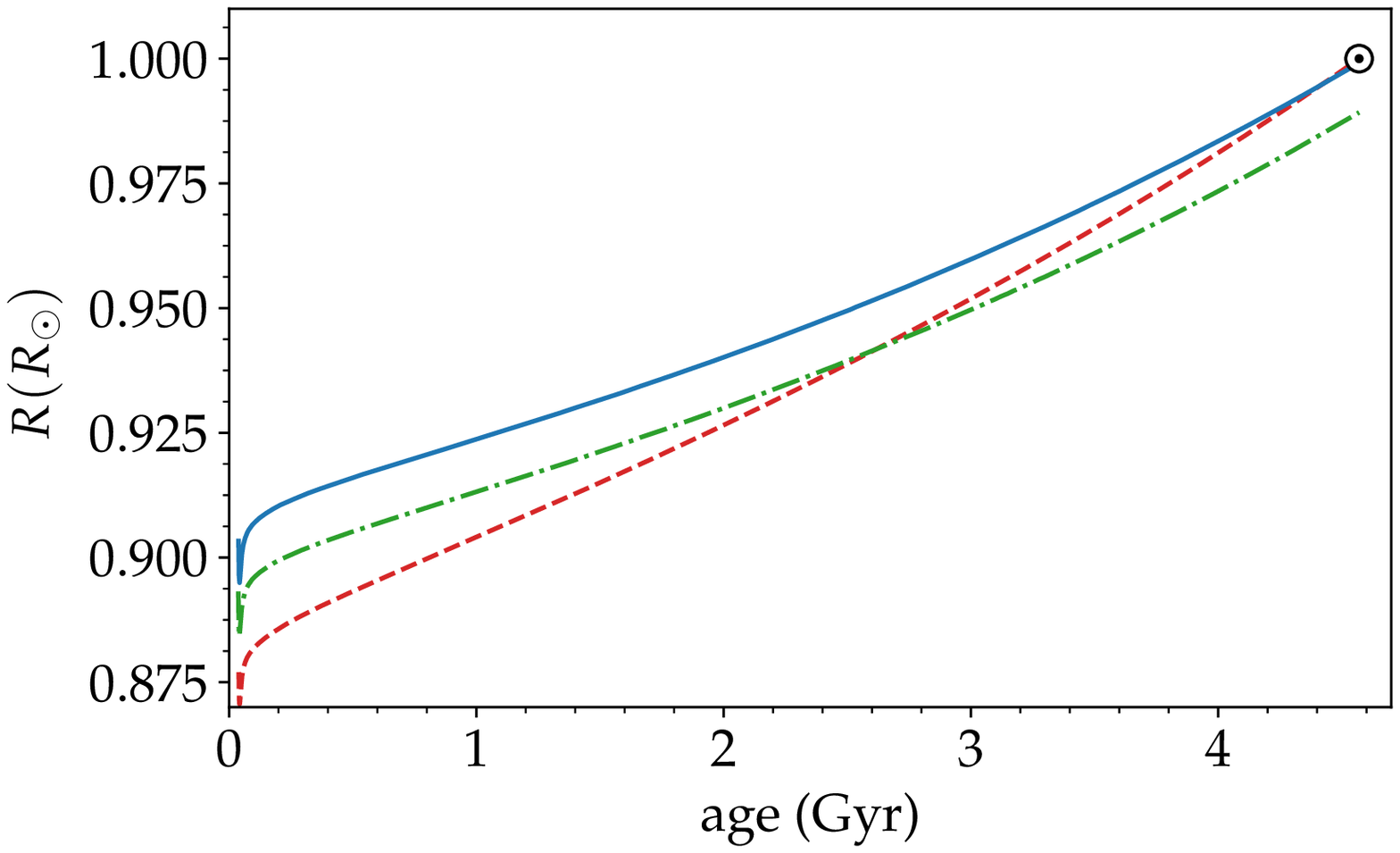}
\caption{Evolution from ZAMS to solar age of the luminosity and the radius for models SSM, ESM0, and ESM1.}
\label{global_from_zams}
\end{center}
\end{figure*}

\subsection{Evolution of the ESM}
\label{evolution}

The entropy-based calibration of \cmlt{} vs. $s_{\rm ad}$ yields an improved evolutionary history of the Sun, in particular for those parameters, such as the surface radius and the depth of the convection zone, which are sensitive to \cmlt{}.

The evolution from ZAMS until solar age of the luminosity and the radius is plotted in Figure \ref{global_from_zams}.
The luminosity (left panel) is remarkably unaffected by the calibration of \cmlt{}, resulting in almost indistinguishable luminosity tracks for the entropy- and the solar-calibrated models.
This is not surprising, since the total luminosity output of a star is essentially controlled by the opacity of its outer layers, i.e., by their ability to sustain an energy flux in equilibrium with the energy production in the interior.  
The opacity is only affected indirectly by the revised \cmlt{} calibration, as a consequence of the change in the metallicity of the convective envelope due to the different history of the depth of the convection zone (see Section \ref{interior}).
Such a small, second order effect has a negligible impact on the luminosity.
This result is consistent with the modest differences found in the interior structure at solar age, discussed in the previous section.

The most significant differences between the solar- and entropy-calibrated evolution are observed in the radius and in the properties of the convection zone. 
The evolution of the radius (right panel of Figure \ref{global_from_zams}) mimics those of $s_{\rm ad}$ and $d_{\rm CZ}$ shown in Figure \ref{entropy}. 
For these three quantities, the difference between the entropy- and the solar-calibrated tracks decreases monotonically from a maximum value at the ZAMS ($\Delta R/R \approx 2.8 \%$; $\Delta s_{\rm ad}/s_{\rm ad}\approx 1.3 \%$; $\Delta d_{\rm CZ}/d_{\rm CZ}\approx -1.0 \%$) to zero (by construction) at solar age.

It is unclear what effect a small difference in the radius and mean density of the Sun might have had on the early evolution of the solar system.  
Objects that are sufficiently close to the Sun are irradiated by a luminous disk rather than by a point source of light, and this effect could be relevant, for example, for the history of the ice caps at the lunar poles \citep{Siegler_ea:2016}.  
Moreover, the exposure to a lower effective temperature during several Gyrs may have significant consequences for the proto-Earth and other inner solar system objects. 
Modeling the evolution of the atmosphere of the proto-Earth realistically should take into account the knowledge of the spectral history, besides that of the intensity, of the solar radiation, and not be assumed to depend simply on the total solar luminosity.  

\begin{figure}
\begin{center}
\includegraphics[width=0.49\textwidth]{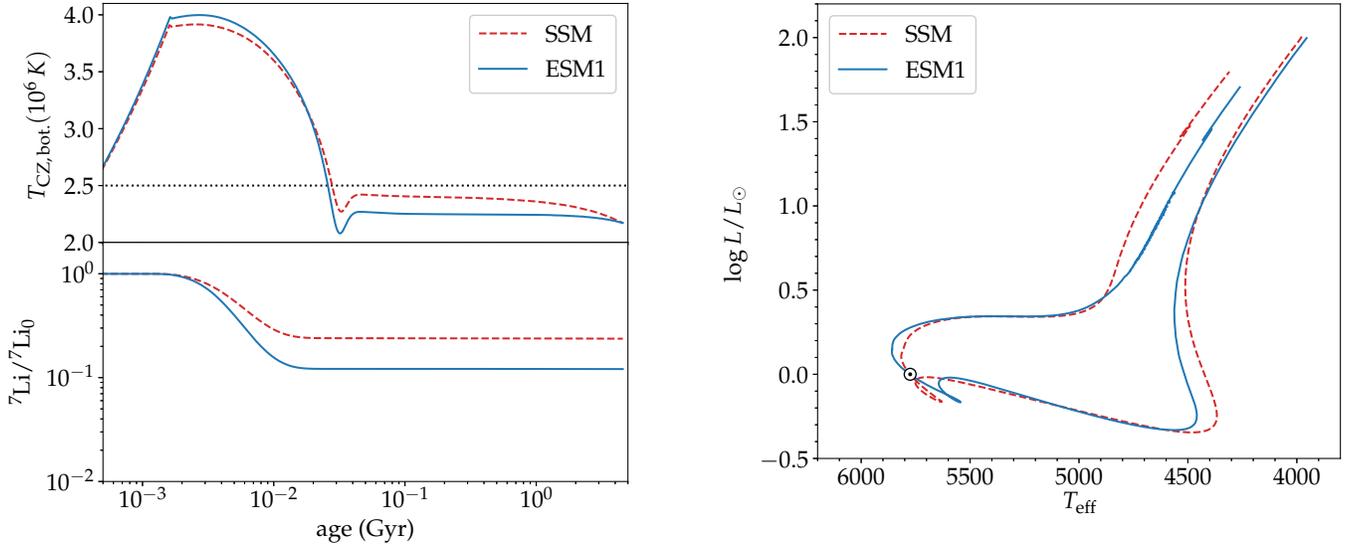}
\caption{Upper panel: temperature at the lower boundary of the convection zone in the entropy-calibrated vs. standard track. The black dotted line marks the threshold for efficient lithium destruction, at $2.5\cdot 10^6$ K. Lower panel: surface lithium abundance normalized to its initial value ($2.0\cdot 10^{-9}$ in terms of mass fraction) for the same models.}
\label{lithium}
\end{center}
\end{figure}

Another significant consequence of the entropy-based calibration is the impact on the depth of the convection zone, which affects the depletion of light elements in the solar atmospheres \citep[see][for a review]{Piau:2001}.  
Since, e.g., $^7$Li is efficiently destroyed at temperatures $\gtrsim 2.5 \cdot 10^6$ K \citep[e.g.][]{C68}, a small change in the temperature at the lower boundary of the convection zone can affect the surface lithium abundance significantly.
As Figure \ref{lithium} shows, the entropy calibration produces a higher temperature at the bottom of the convection zone during the pre-main sequence phase (upper panel), resulting in an increased lithium depletion (approximately by a factor of $2$; cf. lower panel) with respect to the solar-calibrated evolution.
For comparison, \citet{Salaris_Cassisi:2015} report that the use of their \cmlt{} and $T$--$\tau$ relation calibrated on 3D RHD simulations results in a change of the pre-main sequence lithium depletion of the order of $10\%$ with respect to their reference calculations.

No significant lithium destruction occurs in either model during the main sequence. 
Indeed, the observed depletion of $^7$Li (by a factor of about $200$ in the Sun, compared to the primoridial abundance, assumed to be equal to the meteoritic one) cannot be explained in solar-like stars without introducing ad-hoc additional mixing below the convection zone (by a different amount in each case), e.g., for the Sun and the solar analogs HD96423 and  HD38277 \citep{Carlos_ea:2016} and 16 Cyg A \citep{Ramirez_ea:2011}.
This is a long-standing issue of standard stellar models \citep[see, e.g.,][]{Pinsonneault:1997, Somers_Pinsonneault:2014}, and an extensive treatment of the subject is beyond the scope of the present paper.

\begin{figure}
\begin{center}
\includegraphics[width=0.49\textwidth]{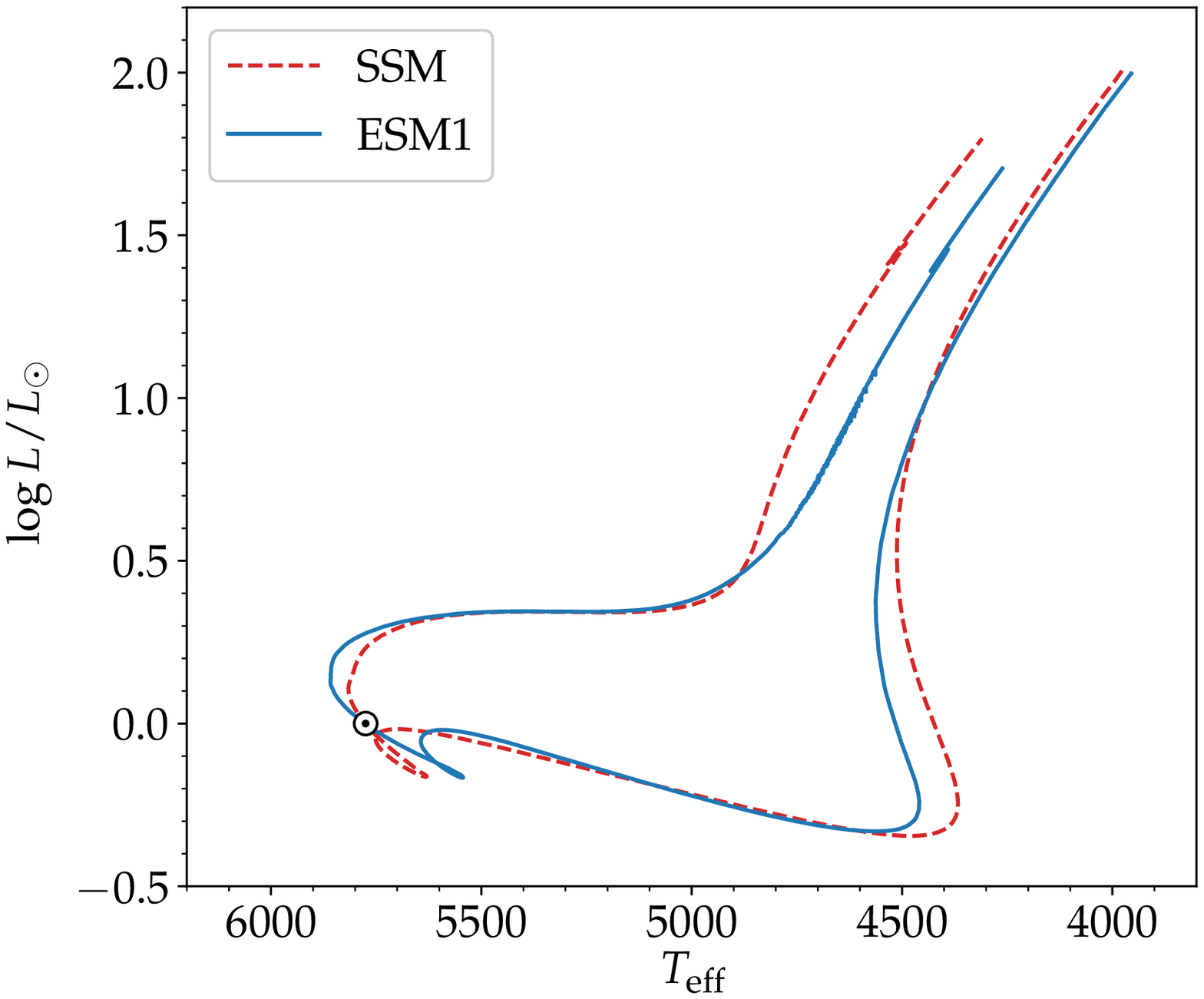}
\includegraphics[width=0.49\textwidth]{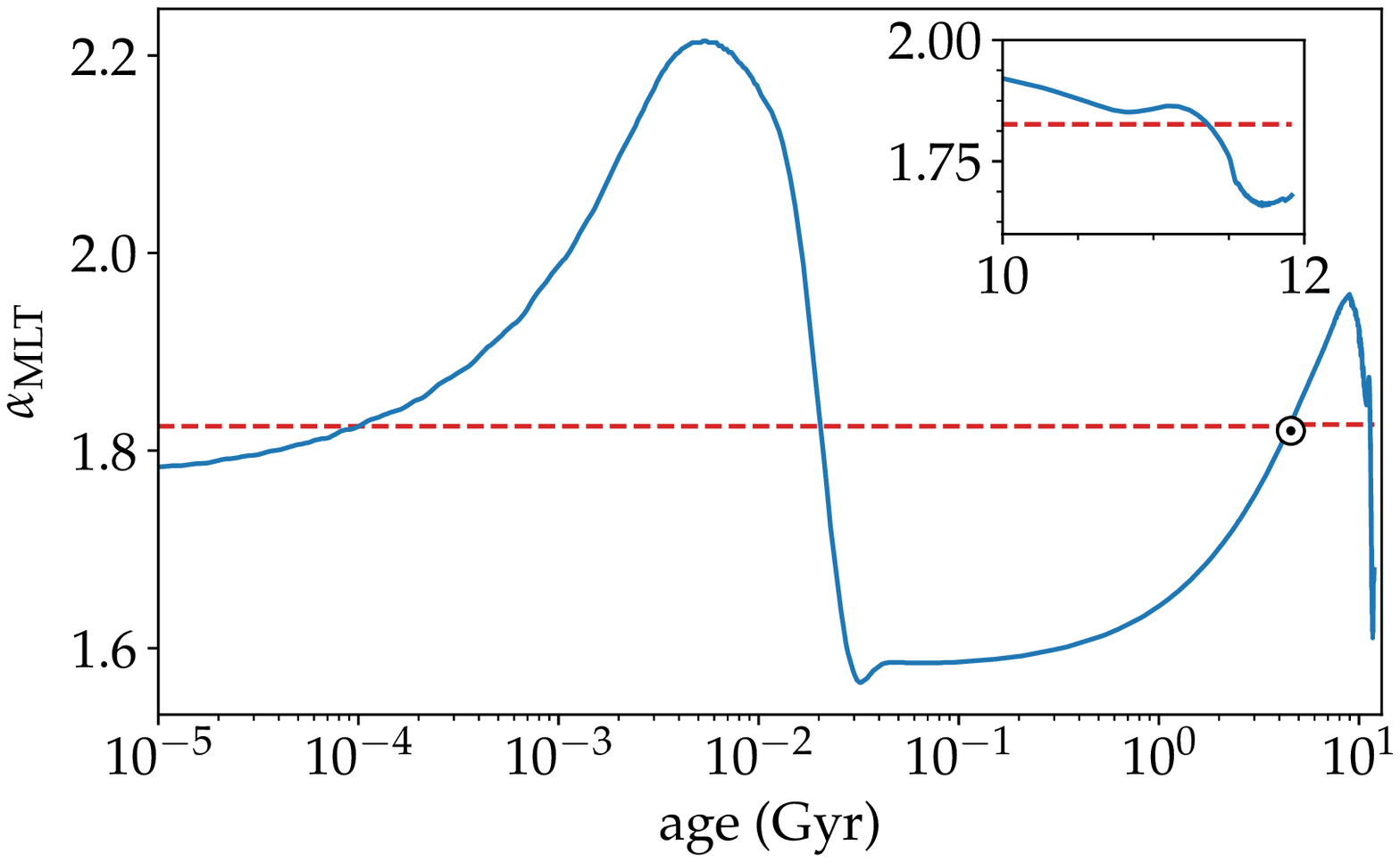}
\caption{Upper panel: evolution in the HR-diagram of the entropy-calibrated and the solar-calibrated tracks. Lower panel: evolution of the MLT parameter for the entropy-calibrated track compared with the (constant) solar-calibrated value. The inset plot shows the red giant branch phase in detail.}
\label{long_evol}
\end{center}
\end{figure}

The complete evolution in the HR diagram from the pre-main sequence to the early red giant phase for the entropy-calibrated and the solar-calibrated tracks is shown in Figure \ref{long_evol}.
The largest differences between the two tracks are found during the pre-main sequence, near the transition from the Hayashi to the Henyey portions of the track, and during the red giant phase.
The approach to the ZAMS and the shape of the main sequence turn-off are also affected.

The evolution of the entropy-calibrated \cmlt{} is plotted in the lower panel of Figure \ref{long_evol}.
Interestingly, the entropy-calibrated \cmlt{} intersects the constant solar-calibrated value several times during the evolution (including, by construction, at solar age).
After the onset of the RGB phase, the entropy-calibrated MLT parameter settle to an approximately constant value, $\alpha_{\rm MLT}\approx 1.65$, lower than the solar-calibrated one.

\section{Discussion}
\label{discussion}

We have presented a first implementation of the method, proposed by \citet{Tanner_ea:2016}, to incorporate the results of 3D RHD simulations of convective envelopes in a 1D stellar evolution code.
This is an alternative approach to the standard calibration of the value of the MLT parameter on the known parameters of the Sun.

A major advantage of our method is the replacement of the MLT parameter \cmlt{} with the value of the adiabatic specific entropy of the convection zone $s_{\rm ad}$ as the main quantity used in the calibration. 
This represents a shift from a purely numerical parameter, which does not have a meaningful interpretation that is not tied to a specific implementation of MLT, to a well-defined physical quantity, independent of the details, and even of the usage of MLT itself to describe convection.  
This approach sets the stellar radii of late-type stars predicted by stellar evolution models on firmer physical grounds.

Along our evolutionary tracks, all models are consistent with the adiabatic specific entropy predicted by the 3D RHD simulations of  \citet{Magic_ea:2013a,Magic_ea:2013b,Magic_ea:2015a,Magic_ea:2015b} and \citet{Tanner_ea:2013a,Tanner_ea:2013b,Tanner_ea:2014}, parameterized as a function of $T_{\rm eff}$, $\log g$, and surface metallicity according to the formulation of \citet{Tanner_ea:2016}. 
The details of the implementation in the Yale stellar evolution code YREC are described in Section \ref{methods}.

In comparing our approach with other alternative methods discussed in the literature, it should be noted that it is not possible to implement realistically the results of 3D simulations into a 1D stellar evolution code without some loss of information in exchange for the simplicity of the treatment in the 1D stellar model. 
Simplifying approximations must be made, in particular, in devising the mapping process.  
With this in mind, it is instructive to compare our implementation with other alternative techniques that have been proposed in the literature.
 
The earliest attempts at a calibration of the MLT parameter based on simulations of convection were based on extracting an effective \cmlt{} by comparing the average profiles obtained from the simulations with 1D stellar envelopes constructed with standard MLT (e.g., \citealt{Ludwig_ea:1999}, using 2D RHD simulations; \citealt{Trampedach_ea:2014a, Trampedach_ea:2014b}, using 3D RHD simulations).

\citet{Salaris_Cassisi:2015} constructed for the first time evolutionary tracks for stars between $0.75$ and $3.0\, M_\odot$ using a 1D stellar evolution code implementing opacities, $T$-$\tau$ relation, and (variable) \cmlt{} consistent with the 3D RHD simulations of \citet{Trampedach_ea:2014a,Trampedach_ea:2014b}.
According to their results, the $T$-$\tau$ relation extracted from the 3D RHD simulations has the strongest effect on the effective temperature scale of the models (up to $\approx 100$ K), while the variable \cmlt{} has a more moderate impact ($30$--$50$ K).

The next step in increasing order of complexity is to replace altogether the outer layers of a 1D stellar interior model with the envelope profile from the RHD simulations, a method usually referred to as ``patching".

The notion of patching was introduced by \citet{Rosenthal_ea:1999} to correct for the surface errors in helioseismic frequencies introduced by the 1D solar model \citep[see also, e.g.,][]{Magic_Weiss:2016}.  
In its original formulation, the patching method is only suitable to improve 1D static models.
This approach has since then been further developed to construct evolutionary sequences (``on-the-fly" patching, \citealt{Jorgensen_ea:2017}).
Patched solar models require a scaling of \cmlt{} that results in a rescaling of the radius to match the solar radius at the solar age. 

Most recently, \citet{Mosumgaard_ea:2018} have implemented  in stellar evolution calculations both the \cmlt{} and the $T$-$\tau$ relation extracted from the 3D RHD simulation by \citet{Trampedach_ea:2014a, Trampedach_ea:2014b}.

All these approaches can be contrasted with our method, in which, as stated above, the key ingredient is the adiabatic specific entropy.  
This difference is of crucial importance. 
Both the patching method and the recalibration of \cmlt{} and of the $T$--$\tau$ relation based on the 3D RHD simulations require a full consistency of the microphysics used in the simulations and in the stellar evolution code \citep[see, e.g., section 4 of][]{Mosumgaard_ea:2018}.
Their main goal is to optimize the agreement of oscillation frequencies to the observed solar frequencies. 
In this paper, on the other hand, we focus on the single objective of predicting reliable radii for late-type stars.  
To this end, what matters for the 1D evolution and its overall radius calibration is only the value of $s_{\rm ad}$. 

\begin{figure}
\begin{center}
\includegraphics[width=0.49\textwidth]{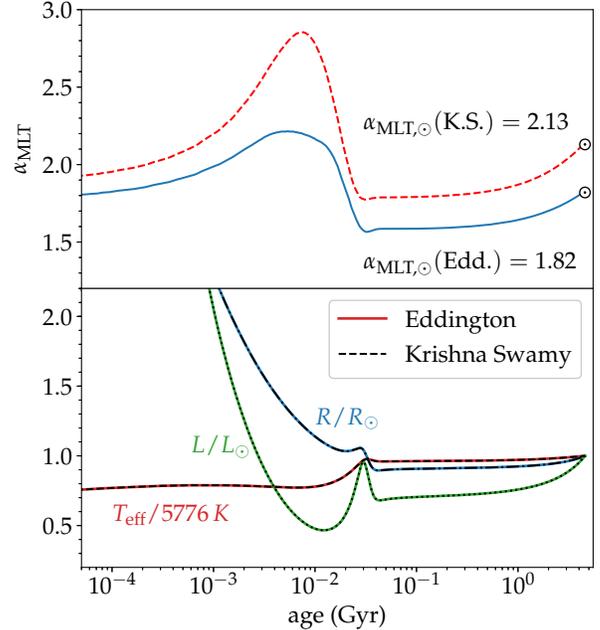}
\caption{Comparison between two entropy-calibrated runs (ESM1) implementing different choices of the $T$--$\tau$ relation. Top panel: the numerical value of the entropy-calibrated \cmlt{} differs between the two cases. 
Bottom: the main stellar structure parameter are identical in the two runs.}
\label{ttau_effect}
\end{center}
\end{figure}

Indeed, the adiabatic specific entropy is decoupled from the detailed treatment of the outer layers and of the atmosphere.
To explore the sensitivity of our entropy-based calibration to this physical ingredient, we have constructed a new ESM1-like run, using the \citet{KrishnaSwamy:1966} $T$--$\tau$ relation instead of the Eddington one (every other parameter being equal).
The results are compared in Figure \ref{ttau_effect}. 
The absolute value of the entropy-calibrated \cmlt{} is different in the two cases, as expected. 
Indeed, the numerical value resulting from a standard solar calibration of \cmlt{} is well-known to be sensitive to the $T$--$\tau$ relation chosen.
However, in our entropy-based calibration, \cmlt{} is adjusted to obtain the same $s_{\rm ad}$ in the model. 
As a result, the depth of the convection zone, the radius of the model, and the other main stellar parameters are effectively decoupled from the $T$--$\tau$ relation. 
This result can be simply understood noting that the $T$--$\tau$ relation mostly affects the SAL, but has hardly any influence on the value of $s_{\rm ad}$.
Our calibration is therefore insensitive to any change of the microphysics (such as the $T$--$\tau$ relation) whose impact is limited to the outer layers of the star, and does not alter the value of the entropy in the adiabatic layers.
In the most extreme case, if a non-MLT description of convection were to be used in a stellar model, a constraint based on $\ssim$ would still be meaningful and applicable.

Additionally, the particular parametrization of $\ssim$ in equation \eqref{jdt_fit} exploits the self-similarity of the properties of convection, expressing $s_{\rm ad}$ as a function of the single variable $x$ as defined in equation \eqref{jdt_x}.
This avoids the extra complexity associated with interpolating the parameters extracted from the 3D RHD simulation in $\log g$, $\log T_{\rm eff})$, and composition \citep[e.g.,][]{Mosumgaard_ea:2018}.

Based on the entropy calibration of MLT, we have successfully constructed a model for the evolution of the Sun, the star for which we have the best available observational constraints. 
Our most significant result for the Sun relates to the greater depth of the convection zone during the evolution toward the main sequence for several billion years, with significant implications for the early evolution of the Earth and the early phases of planetary formation in our solar system.
Note also the associated consequences for the history of light element abundances, as well as a more subtle predicted effects on the chemical composition and the sound speed below the convection zone.

Our entropy-calibrated model ESM1 is in excellent agreement with the SSM in the deep interior of the Sun, and is therefore compatible with the SSM nuclear physics and associated neutrino physics.  

Our method has the potential to allow the construction of stellar evolutionary tracks for late-type stars free of the uncertainty of the mixing length parameter.
Many problems in stellar physics, and indeed most problems in stellar populations, require stellar evolutionary models with physically reliable radii (see, e.g., \citealt{Tanner:PhDT, Tanner_ea:2016}, and references therein).   
In studies of integrated stellar populations in distant systems, this technique promises to provide a more physically justified determination of the relative position of the RGB as a function of metallicity in the HRD than currently available.

The price we pay in this entropy-based calibration is in the details of the atmosphere stratification and dynamics, and the evaluation of the "surface term" in helio- and asteroseismology. 
All these fundamental research topics will have to be addressed on a separate, parallel investigation track.

\section{Conclusions}
\label{conclusions}
 
We have described a novel method to calibrate the free parameter of the mixing-length theory of convection in a 1D stellar evolution code.
Our calibration has two main advantages over the currently standard solar calibration: 1) it relies on a quantity of clear physical significance, the specific entropy at the bottom of the convection zone; 
2) it adapts to the current position of the star in the $(\log T_{\rm eff}, \log g)$ plane, and to its current surface composition.

We have constructed an entropy-calibrated solar model, whose parameters match those of the present Sun as closely as one constructed using the traditional calibration approach.
The entropy-based and standard-calibrated models differ very little in the properties of their deep interiors. 
Moderate differences are found in the composition below the bottom of the convection zone.

The entropy calibration of the MLT parameter produces a revised evolutionary history of the Sun.
The evolution of the luminosity is nearly unaffected.
On the contrary, our entropy calibration has a significant impact on the radius of the model, its effective temperature, and the depth of the convection zone.

These results have potentially significant implications for the depletion of light elements in solar-like stars, and for the early evolution and habitability of solar system objects.
In addition, the position of the red giant branch for an evolved solar track, shifted toward lower effective temperatures, is potentially relevant for the color-magnitude diagrams of stellar clusters.

\acknowledgements{FS was supported by the Max Planck Society grant ``Science Projects in Preparation for the PLATO Mission".
SB acknowledges support from NSF grant AST-1514676 and NASA grant NNX16AI09G.}

\appendix

\begin{figure}
\begin{center}
\includegraphics[width=0.49\textwidth]{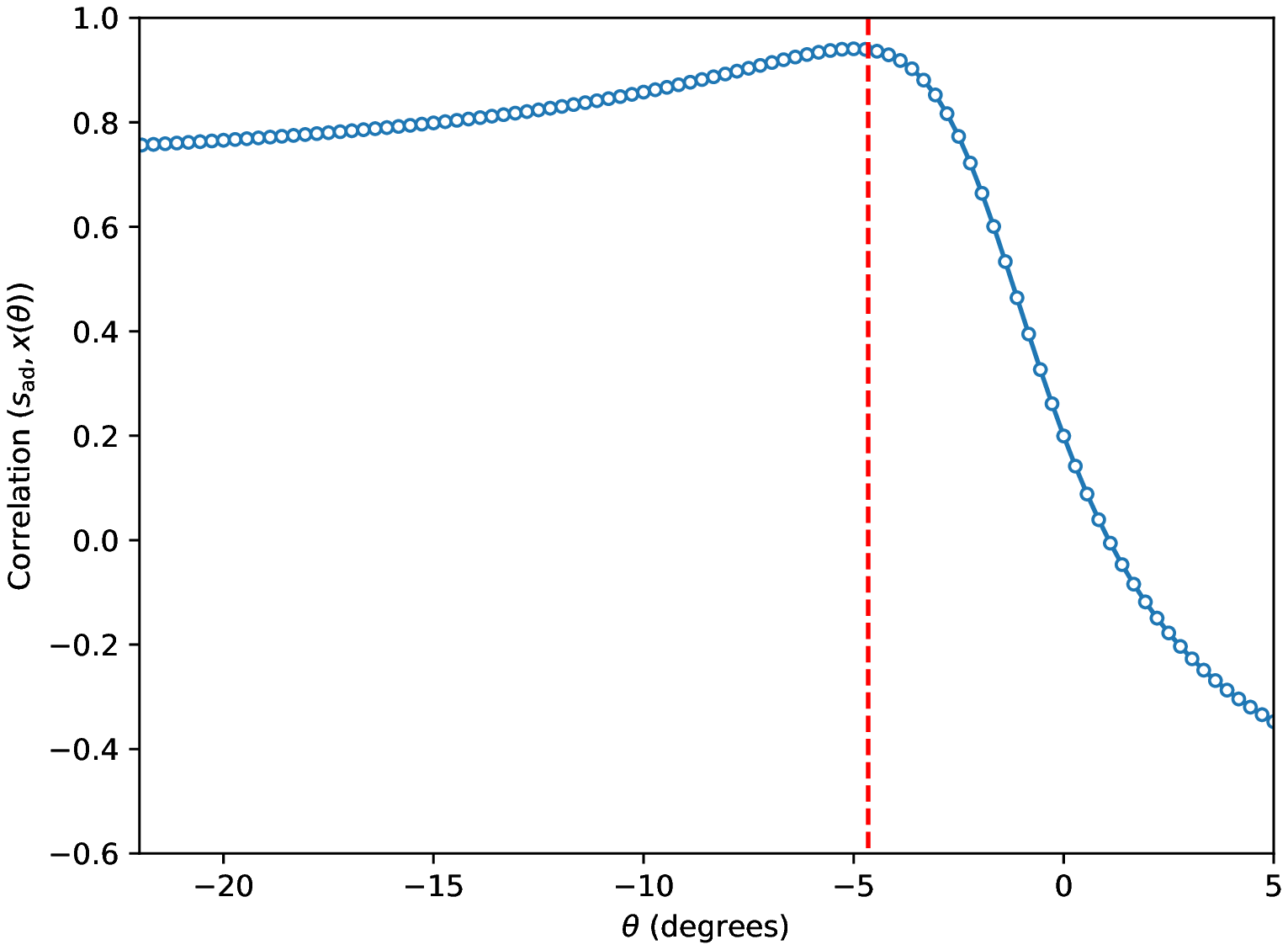}
\includegraphics[width=0.49\textwidth]{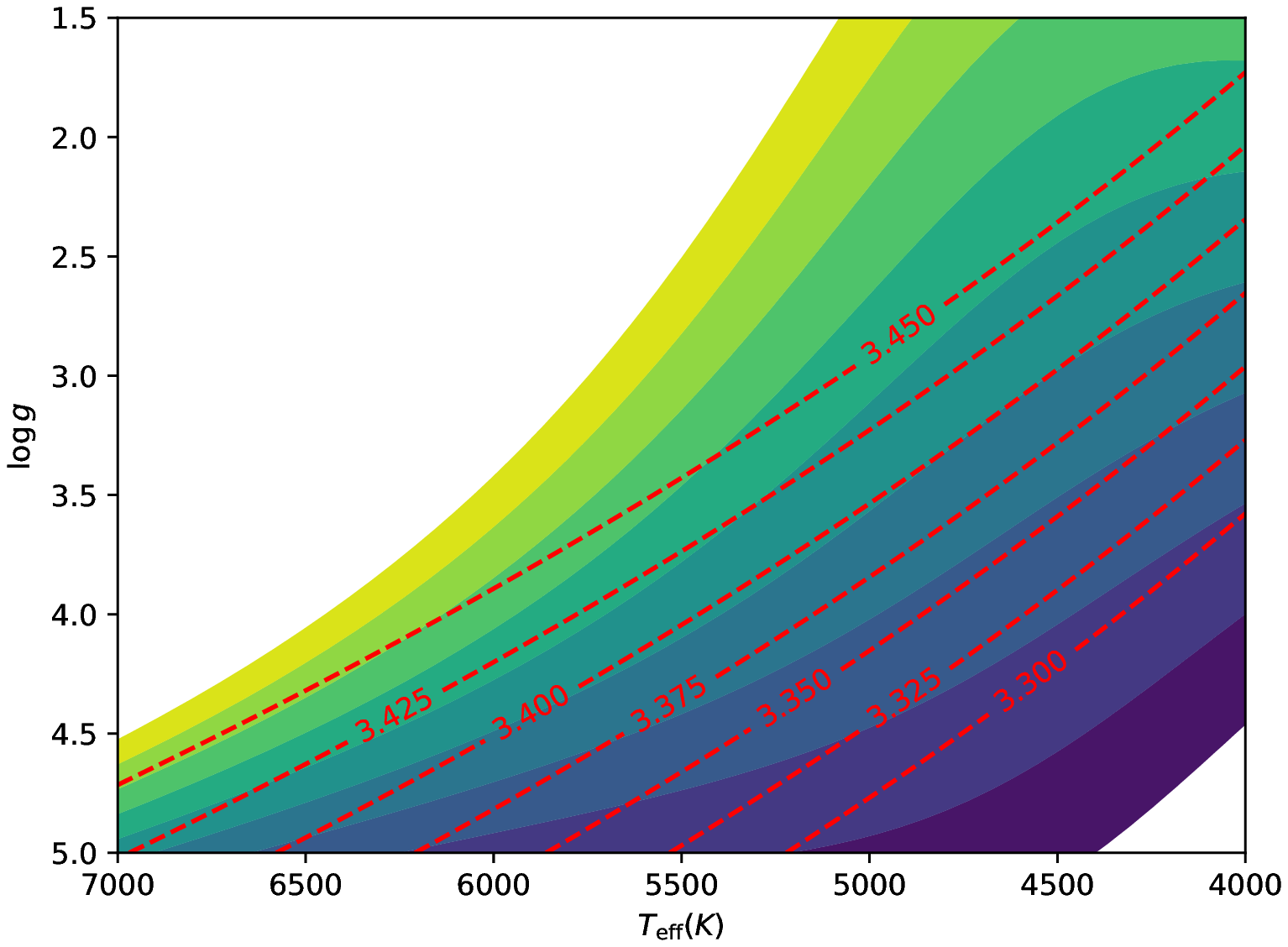}
\caption{Optimal choice of the parameters $A$ and $B$ in equation \eqref{jdt_x} at solar metallicity. Left panel: correlation between $\ssim$ and $x$ as a function of the choice of $A$, $B$, represented as a rotation by the angle $\theta$ ($A=\cos \theta$, $B=\sin\theta$). The vertical line marks the value of $\theta$ corresponding to the $A$, $B$ choice of \citet{Tanner_ea:2016}. 
Right panel: lines of constant $x$ for the \citet{Tanner_ea:2016} solar metallicity $A$, $B$ coefficients (red, dashed), compared with the colored contours of $\ssim$ from \citet{Magic_ea:2015a}.}
\label{correlation}
\end{center}
\end{figure}

\section{Accuracy of the best-fitting formulae for $s_{\rm ad, sims.}$}
\label{app_fits}

The analytical formulae prescribing the mapping between $\{ {\rm [Fe/H]}, T_{\rm eff}, \log g \}$ and $\ssim{}$ play a crucial role in our entropy-based calibration, and it is therefore important to assess the accuracy of their representation of the actual results of the \texttt{Stagger} grid \citep{Magic_ea:2015a} and of the \citet{Tanner_ea:2013a} simulations.
To this end, we have performed an independent re-analysis of the determination of the best-fitting coefficients in equations \eqref{jdt_x} and \eqref{jdt_fit} for the solar metallicity ([Fe/H]$=0$) case.
In this Section, the values of the parameters chosen by \citet{Tanner_ea:2016} will be denoted with the superscript ``T".

The optimal choice of the coefficients $A$ and $B$ results in values of $x$ that are constant along the contours of $\ssim$ in the $(T_{\rm eff}, \log g)$ Kiel diagram; a quantitative measure of this alignment is the correlation between $x$ and $\ssim$.
Since equation \eqref{jdt_fit} is a rotation in the $(T_{\rm eff}, \log g)$ plane, a particular choice of $A$ and $B$ can be represented by a single parameter (the rotation angle $\theta$) as: ($A=\cos\theta$, $B=\sin\theta$).
As shown in the left panel of Figure \ref{correlation}, the original choice $A^T$, $B^T$ coincides with the one producing the maximum correlation of $x$ and $\ssim$. 
The right panel of the Figure shows the comparison of the resulting $x$ with the contours of $\ssim$ from \citet{Magic_ea:2015a}.  
In the range $x=3.30$--$3.45$, the alignment of the lines of constant $x$ with the $\ssim$ contours is satisfactory (cf. Figures 1 and 2 of \citealt{Tanner_ea:2016}).
In the rest of the analysis, we will therefore set the parameters $A$ and $B$ equal to $A^T$ and $B^T$, respectively.

To re-determine the best-fitting values of $s_0$, $\beta$, $x_0$, and $\tau$ we have used a MonteCarlo Markov Chain (MCMC) approach, as implemented in the \texttt{emcee} Python tool \citep{FM_ea:2013}.
We define the logarithm of the likelihood function as follows:
\begin{equation*}
\ln L = -\frac{1}{2} \sum_i \frac{(y_{\rm model}(x_i)-y_i)^2}{\sigma_i^2},
\end{equation*}
where $x_i = A^T \, \log T_{\rm eff,i} + B^T \, \log g_i$, $y_i = s_{\rm ad,i}$, and $\sigma_i = s_{\rm ad, i}^{\rm rms}$ correspond to the solar metallicity subset of the simulations in Table A.1 of \citet{Magic_ea:2015a}, and $y_{\rm model}(x)$ is  given by equation \eqref{jdt_fit}.

We note that the functional form of equation \ref{jdt_fit} contains one parameter, $\beta$, that can be eliminated by an opportune change of variables:
\begin{eqnarray*}
s_{\rm ad} &=& s_0 + \beta \exp\left( \frac{x-x_0}{\tau}  \right) = s_0 + \exp (\ln \beta) \, \exp\left( \frac{x-x_0}{\tau}  \right) =  s_0 + \exp \left( \frac{x - x_0 + \tau \ln \beta}{\tau} \right) = s_0 + \exp\left( \frac{x-x_0'}{\tau}  \right),
\end{eqnarray*}
with $x_0' = x_0 - \tau \ln \beta$.
Doing so reduces the size of the parameter space to be sampled, and minimizes the effect of the correlations among the parameters, thus improving the efficiency of the MCMC procedure.

\begin{figure}
\begin{center}
\includegraphics[width=0.82\textwidth]{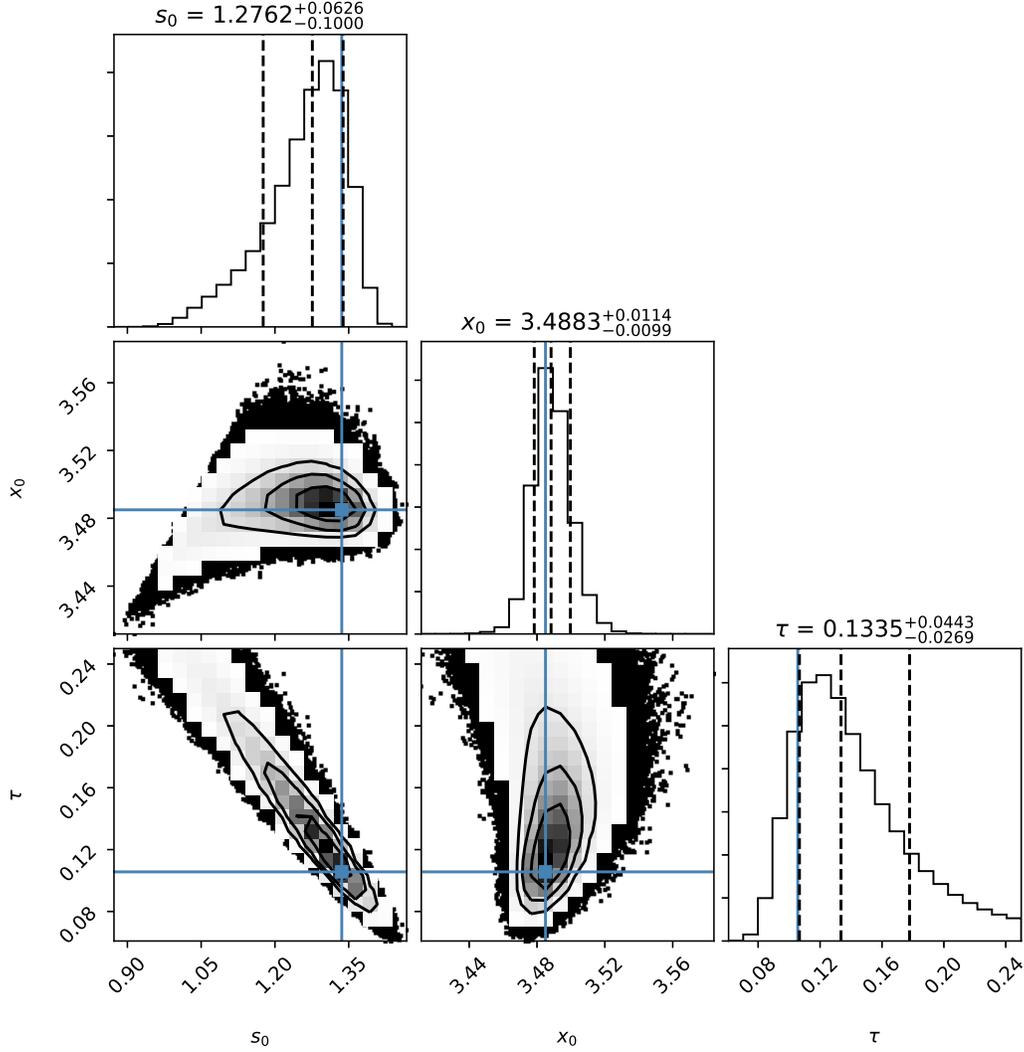}
\caption{
MCMC sampling of the posterior probability distribution for the best-fitting parameters $s_0$, $x_0$, $\tau$ in equation \eqref{jdt_fit} obtained with the \texttt{emcee} package (solar metallicity; $A^T=0.9967$, $B^T=-0.0811$, $\beta^T=1.051$; the length of the chain is $500000$ steps). The solid blue lines mark the original values given in Table 1 of \citet{Tanner_ea:2016}. The dashed lines indicate the locations of the 16th, 50th, and 84th percentiles of the sampled distributions.
}
\label{emcee}
\end{center}
\end{figure}

The MCMC sampling was thus performed for the three parameters $(s_0, x_0, \tau)$, with $\beta=\beta^T$. 
The sampling chain was constructed using $250$ walkers performing $20100$ steps each (the first $100$ steps of each chain were discarded). 
The results are plotted in Figure \ref{emcee}.

In general, all the parameters are in very good agreement with the \citet{Tanner_ea:2016} values. 
By sampling the posterior probability distribution, the MCMC procedure also effectively estimates the uncertainties of the parameters; these are found to be of the order of a few percent, or less. 

Inserting these new values of the parameters in equation \eqref{jdt_fit}, we obtain $s_{\rm ad}\approx 1.776 \cdot 10^{9} \; {\rm erg}\, {\rm g}^{-1}\, {\rm K}^{-1}$, remarkably close to the SSM value (however, this agreement could be coincidental).
In any case, we can conclude that the uncertainty on $\ssim$ due to that on the best-fitting parameters appearing in equation \eqref{jdt_fit} is of the same order of the offset correction introduced in constructing the model ESM1.

\begin{figure}
\begin{center}
\includegraphics[width=0.49\textwidth]{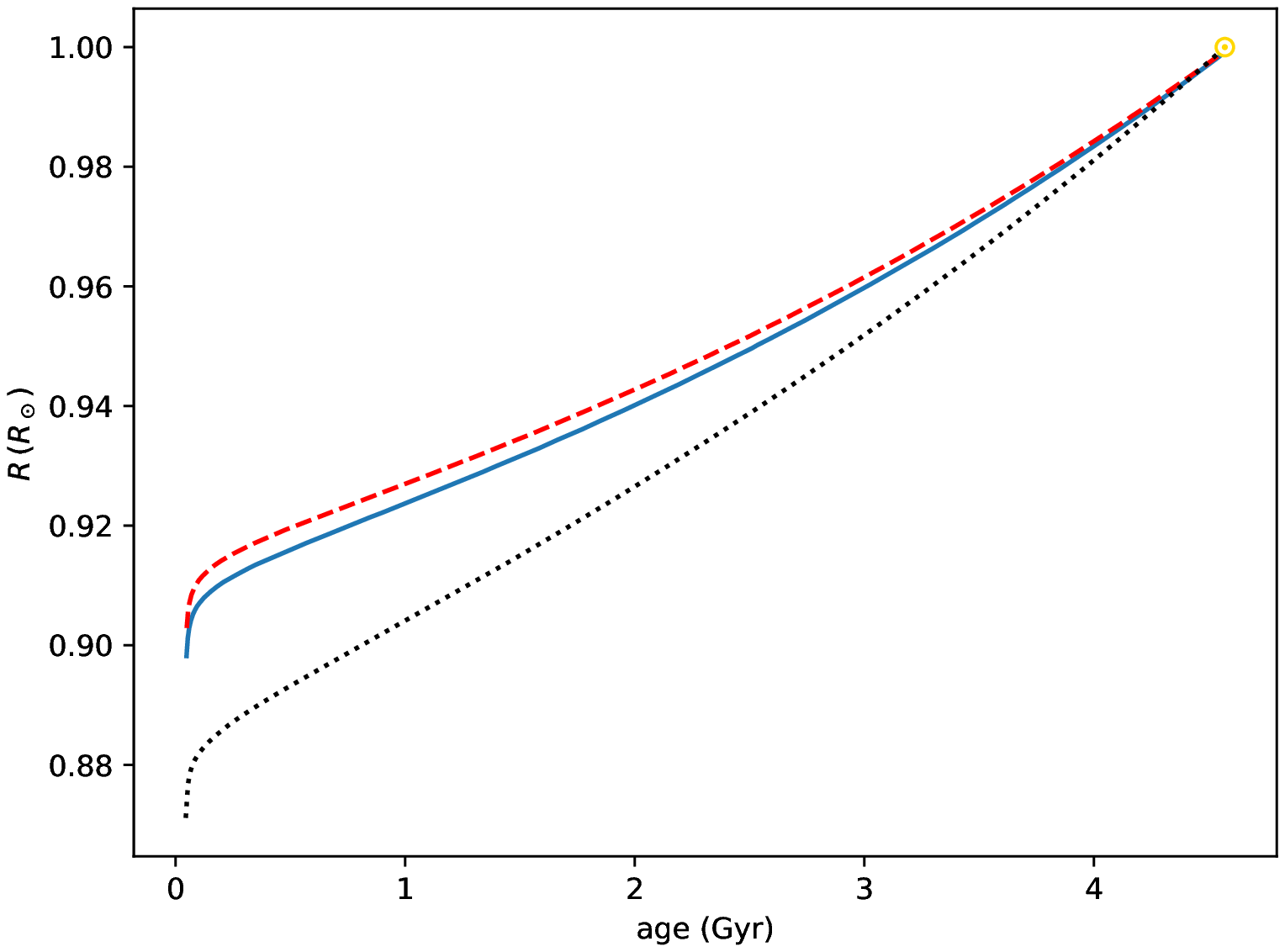}
\includegraphics[width=0.49\textwidth]{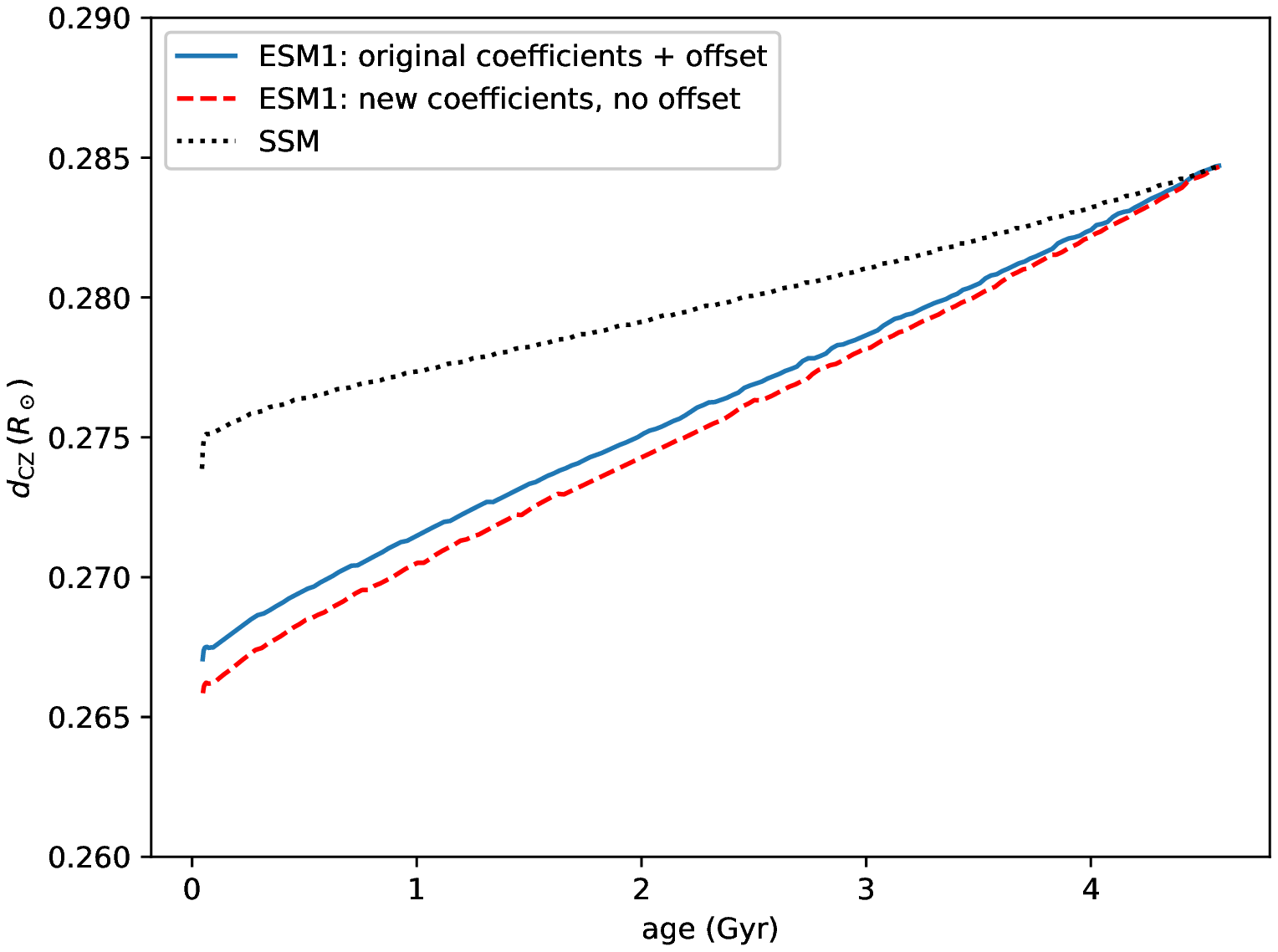}
\includegraphics[width=0.49\textwidth]{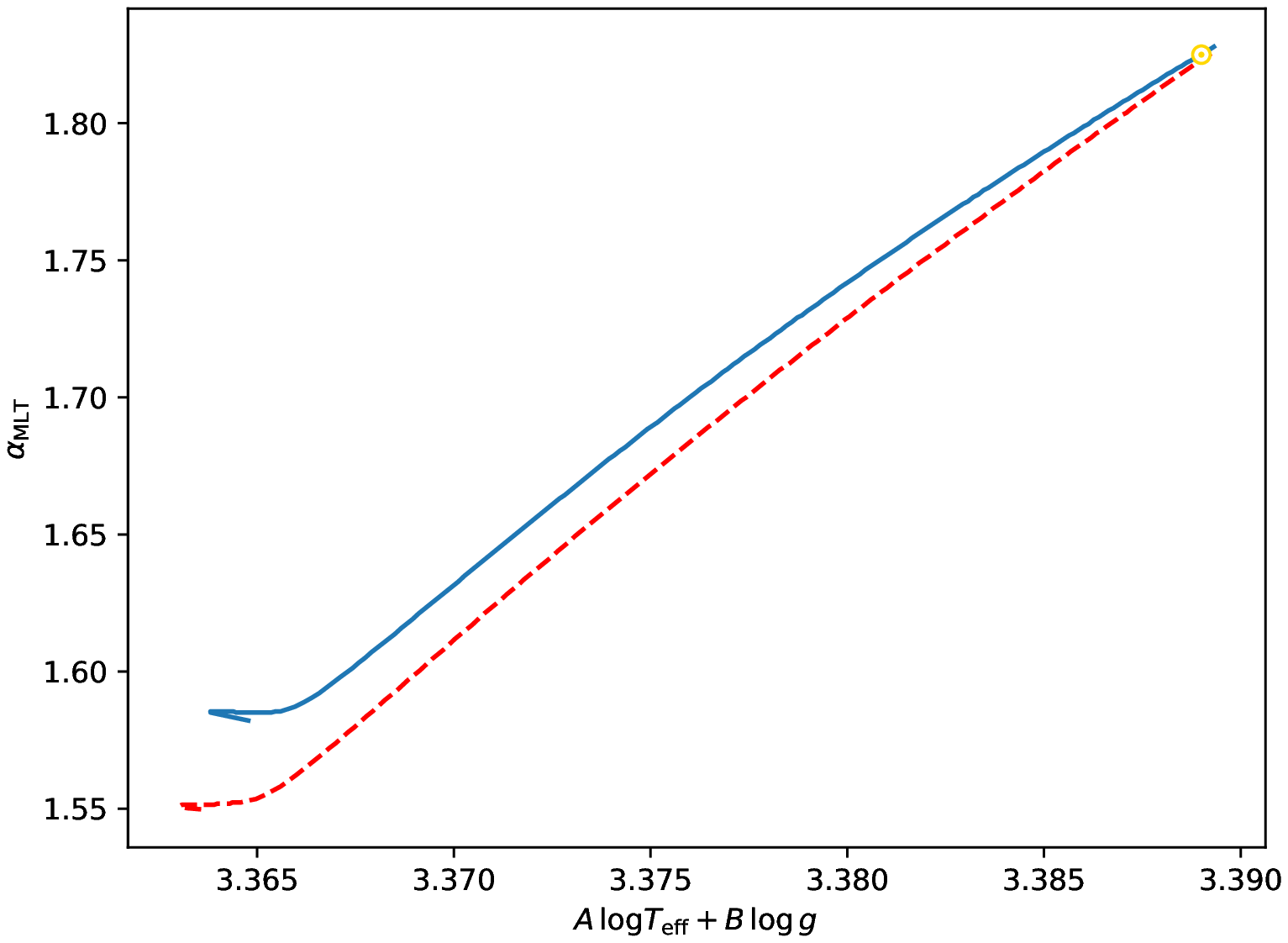}
\includegraphics[width=0.49\textwidth]{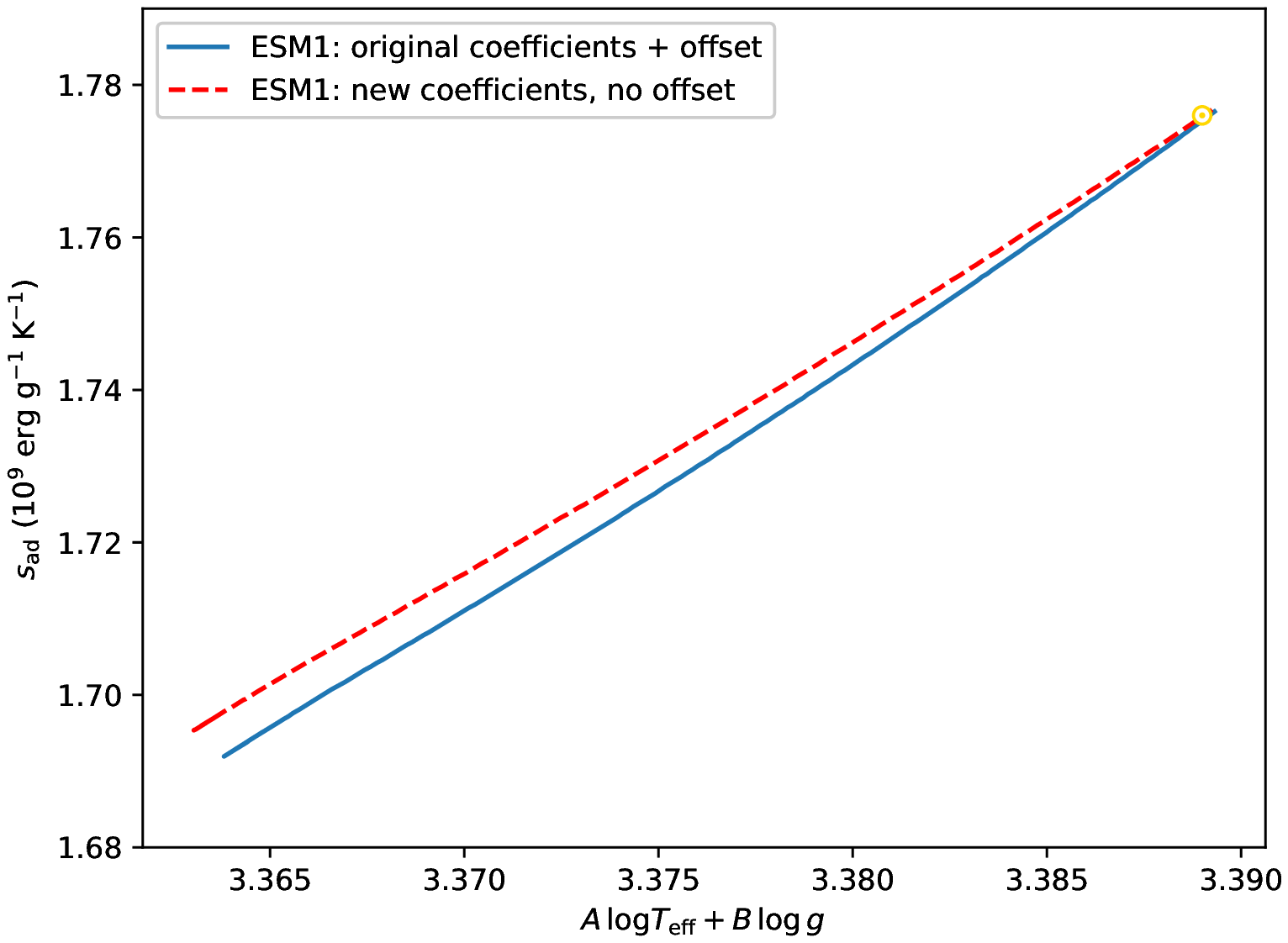}
\caption{Comparison of the entropy-calibrated evolutionary track obtained with the original \citet{Tanner_ea:2016} prescription of the paramters in equation \eqref{jdt_fit} and with their values obtained from our independent MCMC analysis.  
Top row: Evolution of the radius (left) and of the depth of the convection zone (right); the SSM track is also shown for comparison.
Bottom row: \cmlt{} vs. $x$ and $s_{\rm ad}$ vs. $x$ relations (left and right panel, respectively). 
}
\label{newfit}
\end{center}
\end{figure}

To test the impact of relations \eqref{jdt_x} and \eqref{jdt_fit} on the entropy-calibrated evolutionary tracks, we have repeated our ESM1 calculations using the newly determined values of the parameters $s_0$, $x_0$, and $\tau$, and with $\delta s_{\rm ad}=0$.
The results are compared with the original ESM1 discussed in Section \ref{results} and with the SSM in Figure \ref{newfit}.
As can be seen from the top row panels of the Figure, the uncertainty on the best-fitting parameters results in variations in the evolution of the radius and of the depth of the convection zone that are significantly smaller in comparison with the entropy-based vs. solar calibration.
The \cmlt{} vs. $x$ and $s_{\rm ad}$ vs. $x$ relations are also very similar.

Note that $s_{\rm ad}$, \cmlt{}, and the radius have the correct values at solar age both when using the original parameters in equation \eqref{jdt_fit} and the $\delta s_{\rm ad}$ offset, and when using the re-derived parameters without entropy offset. 

We conclude that our results are robust to moderate changes in the details of the representation of $\ssim$ within the range allowed by the uncertainties in the results of the 3D RHD simulations.

\end{document}